\documentclass[useAMS,usenatbib,usegraphicx]{mn2e}

\usepackage{amsmath,amssymb,amsbsy,amsfonts}
\usepackage{mathrsfs}
\usepackage[british]{babel}
\usepackage{mdwmath}
\usepackage{color}
\usepackage{url}

\title{Analytic solutions to the accretion of a rotating finite cloud towards
a central object I.  Newtonian approach.}
\author[S. Mendoza, E. Tejeda, \& E. Nagel]
       {S. Mendoza$^1$, E. Tejeda$^1$ \& E. Nagel$^2$\thanks{Now at 
       Departamento de Astronom\'{\i}a, Universidad de Guanajuato, Guanajuato,
       M\'exico}\\
               $^1$Instituto de Astronom\'{\i}a, Universidad Nacional
           Aut\'onoma de M\'exico, AP 70-264, Distrito Federal 04510,
           M\'exico\\
               $^2$Centro de Radioastronom\'{\i}a y Astrof\'{\i}sica, 
           Universidad Nacional Aut\'onoma de M\'exico, AP 3-72 (Xangari), 
           Morelia 58089, Michoac\'an, M\'exico.
           }

\pagerange{\pageref{firstpage}--\pageref{lastpage}}

\begin{document}

\maketitle
\label{firstpage}
\begin{abstract}
  We construct a steady analytic accretion flow model for a finite
rotating gas cloud that accretes material to a central gravitational
object. The pressure gradients of the flow are considered to be
negligible and so, the flow is ballistic.  We also assume a steady flow
and consider the particles at the boundary of the spherical cloud to be
rotating as a rigid body, with a fixed amount of inwards radial velocity.
This represents a generalisation to the traditional infinite gas cloud
model described by \citet{ulrich76}.  We show that the streamlines and
density profiles obtained deviate largely  from the ones calculated by
\citeauthor{ulrich76}. The extra freedom in the choice of the parameters on
the model can naturally account for the study of protostars formed in
dense clusters by triggered mechanisms, where a wide variety of external
physical mechanisms determine the boundary conditions.  Also, as expected,
the model predicts the formation of an equatorial accretion disc about
the central object with a radius different from the one calculated by
\citet{ulrich76}.
\end{abstract}


\section{Introduction}
\label{introduction}

  The first model of a spherical symmetric accretion flow towards a
central object was made by \citet{bondi52}.  With time, it has become the
key to understand different accretion processes in the universe (see e.g.
\citet{frank02}), despite the fact that it was created for mathematical
curiosity, rather than for astrophysical applications \citep{bondi05}.

  \citeauthor{bondi52} thought of an infinite gas cloud with a
non--relativistic gravitational central object located at the origin
of coordinates.  He assumed the flow to have reached a steady state.
Under these assumptions \citeauthor{bondi52} was able to integrate the
hydrodynamic equations for this spherical symmetric flow.

  Years later, \citet{ulrich76} modified \citeauthor{bondi52}'s ideas by
assuming all fluid particles at the border of the cloud to have a certain
amount of angular momentum, following a distribution given by a rigid body
rotating about the \( z \) axis.   In his model, \citeauthor{ulrich76}
took no account of pressure effects on the infalling gas.  In other
words, his analysis was ballistic, which is approximately true if
the flow is supersonic  and if heating by radiation and 
viscosity effects are negligible \citep[cf.][]{mendoza04}.

  The boundary conditions in \citeauthor{ulrich76}'s model imply that all
fluid particles move in parabolas whose focus correspond to the origin
of coordinates.  As soon as particles arrive to the equatorial plane
they collide with their symmetric counterparts and stay on it, forming an
equatorial disc of radius \( r_\text{u} \) made of fluid particles which
rotate about the central object.  \citeauthor{ulrich76}'s model is usually
taken as the basic model which predicts the formation of an accretion disc
of particles orbiting a central object due to accretion with rotation.
The solution given by \citet{ulrich76} was extended by Casen \& Moosman 
and is used \citep[see e.g.][]{nagel07,lin90,stahler94} as a 
natural initial condition for simulations dealing with the problem of disc
formation and subsequent evolution.

  Both models, \citet{bondi52}  and \citet{ulrich76} have been
taken to the relativistic regime by assuming the central object
to be a Schwarzschild black hole \citep{michel72,huerta05}.
They have wide applications in high energy phenomena \citep[see for
example][]{lee05,beloborodov01}.

  On many astrophysical situations where Ulrich's model is used,
the usual assumption is that the boundary conditions for an infinite
gas cloud remain valid even when it is applied to the inner core of
a large accretion cloud \citep[see e.g.][]{cassen81,lin90,stahler94}.
This is not necessarily true, because the accretion flow at the boundary
of the inner core has, in general, an inward radial velocity different
from zero\footnote{In fact, as we will see later on, if one builds a
stationary situation with zero inward radial velocity, then it would
be necessary to have an infinite density at the cloud's border.}. Also,
if the angular momentum of particles follows rigid body rotation at
this finite radius then, as we describe in this article, the flow may
largely deviate from Ulrich's solutions, since the specific total energy
(kinetic plus gravitational) is not zero and so, orbits are no longer
parabolic. The relaxation of Ulrich's boundary conditions leads to a more
realistic infall model with applications to a wider variety of accretion
problems. With an appropriate selection of boundary conditions this could
in principle modify the main results of a number of previous works in
the literature dealing with various astrophysical accretion problems.

  In this article we assume a central Newtonian potential produced by the
central star, and a rotating gas cloud with a finite radius \( r_0 \).
Based on this, we construct a cylindrical symmetric flow which also
predicts the existence of an accretion disc about the central object.
As it will be shown, the trajectories are no longer sections of parabolas.
They are general conical sections and once boundary conditions are fixed
they can be of different families for each fluid particle.

  In order to show the relevance of having a finite cloud radius \(
r_0 \)  and a non--zero radial velocity \(v_{r_0} \), note that this
boundary conditions can be added to the standard Ulrich's ones which
are defined by the mass \( M \) of the central star, an accretion rate
\( \dot{M} \) and  a specific angular momentum \( h_0 \) of
the cloud.  In other words, \( r_0 \) and \( v_{r_0} \) characterise
the modifications to Ulrich's~(\citeyear{ulrich76}) model.  The length
$r_{0}$ can be taken as the outer boundary of the inner region of the
cloud, where its self-gravity is neglected, and the main force is given
by the gravity of the star.

  In real astrophysical problems this model can be applied. For example to
NGC 1333 IRAS 4 region where an unusual velocity profile in an infalling
envelope is proposed by \citet{choi04} to explain the observation.
Also, it is relevant in the study of triggering star formation, because
whatever the physical mechanism responsible, a whole set of combinations
for $r_{0}$ and $v_{r_{0}}$ is possible. In this scenario, Ulrich's model
is not applicable.  There are well known mechanism for the triggering,
namely stellar ejecta from winds \citep{foster96} or supernova shells
\citep{melioli06} as well as the ionising radiation from a star,
or a collection of stars,  that produces an ionisation shock front
\citep{bertoldi89,esquivel07}.  Each of these mechanisms sets particular
boundary conditions for the collapse of the cloud to form a star.

The triggering is relevant in clouds with a high density of
potential in sites of star formation. A first generation of stars can
influence the dynamical configuration for the next generation. This
mechanism is highly important when it is realized that the stars are
formed in turbulent clouds \citep{vazquez07} and that a quiescent core
for star formation envisioned by \citet{shu87} is in some dense clusters
difficult to find. A complete picture of these ideas is clearly presented
by \citet{hartmann01}.

A glimpse of the full range of boundary conditions is exemplified by
\citet{hartmann07}. They speculate that the rich Orion nebula cluster
is formed by the collapse of an elliptical rotating sheet of gas with a
density gradient along the major axis. A high density region is formed
when one tip of the cloud collapses preferentially and so,
the already formed protostars \citep{heitsch08}
accrete material with velocities given by the global collapse of the
cloud, independent of the gravity of the collapsed object. Thus, free
choosing of boundary conditions is highly important to study the regions
close to the star.  The resolution of these regions is beginning to be
reached by computer simulations and observations.

  In this context, the ballistic solution given in this paper is
useful to study star and disc formation in very complex environments,
where one has the need to choose boundary conditions freely and
easily. However, an obvious weakness in the model is that at $r_{0}$,
the cloud is rigidly rotating. Thus, in the scales where this solution
is valid, one either needs to invoke a magnetic field at $r>r_0$ that
is responsible to enforce rigid body rotation \citep{mouschovias79},
or the fact that the shell feeding the star-disc system rotates
rigidly during some reasonable period of time, in order to consider
this process as quasi-stationary. However, there are many works in the
literature \citep{kenyon93,jayawardhana01,whitney03} that make such an 
assumption.

  In Sections \ref{velocity} and \ref{particle-number} we calculate the
velocity and density fields of the flow respectively.  At the end of
Section~\ref{particle-number} we recover  \citeauthor{ulrich76}'s
(\citeyear{ulrich76}) model.  Section \ref{streamlines} describes the
behaviour of the flow for its different parameters and discusses its
limitations. Finally, Section \ref{application} deals with realistic
astrophysical situations and the validity of our model on those
scenarios.

\section{The model and its velocity field}
\label{velocity}

  Let us assume that a spherical cloud with radius \( r_0 \) accretes
matter in a steady way towards a central object located at the origin
of coordinates.   All fluid particles located at \( r_0 \) are taken to
follow rigid body motion in the azimuthal direction, i.e. they rotate
about the \( z \) axis of coordinates.  At this position, particles also
have a radial velocity component \( v_{r_0} \). Here and in what follows,
due to the symmetry of the problem we use spherical coordinates \(r,\
\theta,\ \text{and} \ \phi \) for the radial coordinate, and the polar
and azimuthal angles respectively.  The particle number density \( n \)
at \( r_0 \) is taken to be constant with a value \( n_0 \).  To simplify
the problem, we assume the self gravity of the gas cloud to be small as
compared to the gravitational potential produced by the central object.
In addition, if the gradients of pressure are small compared to the
kinetic and potential energies, then all fluid particle trajectories can
be approximated as ballistic.  The total angular momentum of the gas
cloud points in the direction of the \( z \) axis, which contains the
central object.  Note that when \( r_0 \rightarrow \infty \) and \( v_{r_0}
\rightarrow 0 \), the model converges to the accretion problem proposed by
\citet{ulrich76}, which is well described by \citet{mendoza04}.  In fact,
Ulrich's convergence is mathematically proved at the end of
Section~\ref{particle-number}.

  Since the gas is being accreted from \( r_0 \), it's accretion flow rate
per unit mass \( \dot{N} \) at any fixed radial distance \( r \) is
constant and given by

\begin{equation}
  \dot{N}  = 4 \pi r_0^2 n_0 v_{r_0} = \text{const}.
\label{eq01}
\end{equation}

  At \( r_0 \), the distribution of specific angular momentum \( h \) 
is given by \( h = h_0 \sin \theta_0 \), where \( \theta_0 \) is the initial
polar angle of the particle's trajectory.

  Under all the above assumptions, the specific energy \( E \) and the
specific angular momentum \( h \) are constants of motion along each
particular trajectory.  The specific energy \( E \) is given by

\begin{equation}
  E = \frac{ 1 }{ 2 } v_r^2 + \frac{ 1 }{ 2 } \frac{ h^2 }{ r^2 } - 
    \frac{ GM }{ r } = \frac{ 1 }{ 2 } v_{r_0}^2 + \frac{ 1 }{
      2 } \frac{ h_0^2 \sin^2 \theta_0 }{ r_0^2 }-\frac{ GM }{ r_0 }.
\label{eq02}
\end{equation}

\noindent where \(G\) is Newton's gravitational constant. We now introduce 
two dimensionless parameters \( \mu \) and \( \nu \) given by 

\begin{equation}
  \mu^2 := \frac{h_0^2 }{ r_0^2  E_0} = \frac{ r_\text{u}^2 }{ r_0^2 }, 
  \qquad \nu^2 := \frac{ v_{r_0}^2 }{ E_0 }, 
\label{eq03}
\end{equation}

\noindent where \( r_\text{u} = h_0^2/GM \) is the disc's radius in Ulrich's
model and \(E_0 := GM / r_\text{u} \) is the specific gravitational potential
energy of a fluid particle evaluated at \( r_\text{u} \). The quantity \(\mu
\) represents the ratio of Ulrich's disc radius to the original cloud's
radius. So, for example, in order to consider an infinite cloud's radius
we must take the limit \(\mu \rightarrow  0\). On the other hand, \( \nu
\) represents the initial amount of radial velocity measured in units of
the Keplerian velocity \( v_\text{k} := \sqrt{E_0} \), at position \(
r_\text{u} \).  With these parameters, equation~\eqref{eq02} can be written
into dimensionless form as

\begin{equation}
  \varepsilon = \nu^2 + \mu^2 \sin^2 \theta_0 - 2 \mu,
\label{eq04}
\end{equation}

\noindent where the dimensionless energy \( \varepsilon \) is given by \(
\varepsilon := 2 E / E_0 \).

  The trajectory of each fluid particle is contained on a plane
and is given by a conic section. The origin represents
one of the foci of the orbit. At a specific initial position, the
particle is located at \( r_0 \) and its polar and azimuthal angles are
given by \( \theta_0 \) and \( \phi_0 \).  Over the orbit plane, the particle
trajectory is defined by an azimuthal angle \( \varphi \) which at the
initial position has the value \( \varphi_0 \).  The trajectory of a
given particle is defined by the solution to Kepler's problem
\citep{daumech}:

\begin{equation}
  r = \frac{ \sin^2 \theta_0 }{ 1 - e \cos\varphi }.
\label{eq05}
\end{equation}

\noindent In the previous equation and in what follows, unless stated
otherwise, distances are measured in units of \( r_\text{u} \).  
The eccentricity \( e \) of the orbit is given by

\begin{equation}
  e = \sqrt{ 1 + \varepsilon \sin^2 \theta_0 }.
\label{eq06}
\end{equation}

\noindent At the border of the cloud \( r = r_0 = 1/\mu \) and so,
substitution of this  in equation~\eqref{eq05} gives the following
condition for \( \varphi_0 \):

\begin{equation}
  \cos\varphi_0=\frac{ 1 }{ e }( 1 - \mu \sin^2 \theta_0).
\label{eq07}
\end{equation}

  Performing standard spatial rotations its easy to obtain the following
formulae between the angles \(\varphi, \ \varphi_0, \ \theta, \ \theta_0,\
\phi \text{ and } \phi_0 \):

\begin{equation}
\cos( \varphi - \varphi_0 ) = \frac{ \cos\theta }{ \cos\theta_0 }
, \qquad \cos( \phi - \phi_0 ) = \frac{ \tan\theta_0 }{ \tan\theta }.
\label{eq08}
\end{equation}

  Using equation~\eqref{eq08} to rewrite equation~\eqref{eq05} yields

\begin{equation}
  r = \frac{ \sin^2 \theta_0 }{ 1 - e \cos\xi },
\label{eq09}
\end{equation}

\noindent where 

\begin{equation}
  \xi = \cos^{-1} \left( \frac{ \cos\theta }{ \cos\theta_0 }
\right) + \varphi_0.
\label{eq10}
\end{equation}

  We now use the fact that in spherical coordinates \( v_\phi = r \sin
\theta \mathrm{d} \phi / \mathrm{d} t = h \sin \theta_0 / r \sin \theta \),
\( v_\theta = r \mathrm{d} \theta / \mathrm{d} t = \left( \mathrm{d} \theta 
/ \mathrm{d} \phi \right) \left( v_\phi / \sin \theta \right) \), and
\( v_r =  \mathrm{d} r / \mathrm{d} t = \left( \mathrm{d} r / \mathrm{d} 
\theta \right) \left( v_\theta / r \right) \).
With this and using equation~\eqref{eq08} the expressions for the velocity
field, measured in units of \( v_\text{k} \) are given by:

\begin{equation}
v_\phi = \frac{ \sin^2 \theta_0 }{ r \sin\theta },
\label{eq11}
\end{equation}

\begin{equation}
v_\theta = \frac{ \sin \theta_0 }{ r \sin \theta }
\left( \cos^2 \theta_0 - \cos^2 \theta \right)^{ 1/2 },
\label{eq12}
\end{equation}

\begin{equation}
v_r = -\frac{ e \sin \xi \sin \theta_0 }{ r ( 1 - e \cos\xi ) }.
\label{eq13}
\end{equation}

\section{Particle number density}
\label{particle-number}

  In order to obtain the particle number density field we start from
the continuity equation for a stationary flow, i.e.

\begin{equation}
\nabla \cdot ( n \boldsymbol{v} ) = 0.
\label{n1}
\end{equation}

  We now integrate the previous equation over a volume consisting on
a collection of streamlines.  In other words, this ``tube'' is built
in such a way that its lateral surface is bound by streamlines and its
lids by spherical sections, the upper one at the cloud's border, i.e.  \(
r = r_0 = 1/\mu \) and the lower at any arbitrary radial distance \( r \)
such that \( 0 < r < 1 / \mu \).  With this selection, equation~\eqref{n1}
can be integrated to obtain:

\begin{equation}
n \boldsymbol{v} \cdot \mathrm{d} \boldsymbol{a} \bigg |_{r=1/\mu} = 
n \boldsymbol{v} \cdot \mathrm{d} \boldsymbol{a} \bigg |_r.
\label{n2}
\end{equation}

\noindent Taking into account the fact that the differential area element
\( \mathrm{d} \boldsymbol{a} \) is given by

\begin{equation}
\mathrm{d} \boldsymbol{a} = r^2 \sin \theta ~\mathrm{d} \theta 
~\mathrm{d} \phi ~\boldsymbol{e}_r + 
r \sin \theta ~\mathrm{d} r ~\mathrm{d} \phi ~\boldsymbol{e}_\theta
+ r ~\mathrm{d} r~ \mathrm{d} \theta ~\boldsymbol{e}_\phi,
\label{n3}
\end{equation}

\noindent then, equation\eqref{n2} transforms to

\begin{equation}
 \sin \theta_0 
~\mathrm{d} \theta_0 ~\mathrm{d} \phi_0 = -n \, v_r ~r^2 
\sin \theta ~\mathrm{d} \theta ~\mathrm{d} \phi,
\label{n4}
\end{equation}

\noindent where the particle number density \( n \) is measured in units of
\( n_\text{u} := \dot{N} / 4 \pi v_\text{k} r_\text{u}^2 \).
Using the symmetry of the quantities in the azimuthal angle
involved in the previous equation, we can integrate equation~\eqref{n4}
with respect to \( \phi \) from \( 0 \text{ to } 2\pi\), to obtain

\begin{equation}
n = \frac{ 1 }{ r^2 } \frac{ \sin \theta_0 }{ \sin \theta } 
\left[ - v_r 
\left( \frac{ \partial \,\theta }{ \partial \theta_0 } \right)_r
\right ]^{ -1 }.
\label{n5}
\end{equation}

\noindent On the other hand, from equation \eqref{eq09} it follows that

\begin{equation}
  \begin{split} 
 -v_r \frac{ \sin \theta }{ \sin \theta_0 } & \left( \frac{ \partial \,\theta }
 { \partial \theta_0 } \right)_r  = \frac{ 1 }{ \sin \theta_0 }  
    \bigg\{ \bigg( 1 + \frac{ 3 \cos^2 \theta_0 -1 }{ r }
     \ - 
   					  \\	
     & - 2 \mu \cos \theta_0 \cos \theta \bigg) 
      \left( 1 - \frac{\cos^2 \theta}{\cos^2 \theta_0} \right )^{1/2} +
     					\\
     & +  \frac{ \nu }{ \sin \theta_0 } \left ( 1 + \cos^2 \theta - 
       2 \cos^2 \theta_0 \right ) \bigg\}.
  \end{split}
\label{n6a}
\end{equation}

\noindent Substitution of equations~\eqref{n6a} and \eqref{eq13} into 
\eqref{n5} leads to the required particle number density field:

\begin{equation}
  \begin{split}
    n = & \frac{ \sin \theta_0 }{ r^2 } 
    \bigg\{ \bigg( 1 + \frac{ 3 \cos^2 \theta_0 -1 }{ r }
     \ - 
   					  \\	
     & - 2 \mu \cos \theta_0 \cos \theta \bigg) 
       \left( 1 - \frac{\cos^2 \theta}{ \cos^2 \theta_0 } \right )^{1/2} +
     					\\
     & + \frac{ \nu }{ \sin \theta_0 } \left ( 1 + \cos^2 \theta - 
       2 \cos^2 \theta_0 \right ) \bigg\} ^{-1}.
   \end{split}
\label{n6}
\end{equation}

  As a result of the collision at the equatorial plane between the
streamlines coming from the northern hemisphere with those coming from
the southern one, a shock front is expected to form, as well as the
formation of an equatorial disc shape structure \citep[see e.g.][and
references therein]{ulrich76,mendoza04}.

  Taking \( \theta \rightarrow \pi / 2 \) in equation~\eqref{eq09}
followed by the substitution \( \theta_0 \rightarrow \pi / 2 \) we obtain
an expression for the accretion disc radius given by:

\begin{equation}
  r_\text{d} = \frac { 1 }{ 1 + \nu }
\label{d1}
\end{equation}

  From this last equation and since naturally \( r_d <
r_0 \), the following condition between the parameters arises:

\begin{equation}
  \mu < 1 + \nu
\label{eea}
\end{equation}

  If in equations~\eqref{eq09}, \eqref{eq11}, \eqref{eq12}, \eqref{eq13} and 
\eqref{n6} we take the parameters values \(\mu = 0\) and \(\nu=0\) ,
we get the following set of equations

\begin{equation}
  r = \frac{ \sin^2 \theta_0 } { 1 - \left( \cos \theta / \cos\theta_0
    \right)  } ,
\label{u1}
\end{equation}

\begin{equation}
  v_r = -\frac{ \sin \theta_0 }{  r }\left(\frac{ \cos\theta_0 + \cos\theta }   
	{ \cos\theta_0 - \cos\theta } \right)^{1/2} ,
\label{u2}
\end{equation}

\begin{equation}
v_\theta = \frac{ \sin \theta_0 }{ r \sin \theta }
\left( \cos^2 \theta_0 - \cos^2 \theta \right)^{ 1/2 },
\label{u3}
\end{equation}

\begin{equation}
v_\phi = \frac{ \sin^2 \theta_0 }{ r \sin\theta },
\label{u4}
\end{equation}

\begin{equation}
  n = \frac{ \sin \theta_0 }{ r^2 } 
    \bigg( 1 + \frac{ 3 \cos^2 \theta_0 - 1 }{ r }
    \bigg)^{-1} \left( 1 - \frac{ \cos^2 \theta }{ \cos^2 \theta_0 }
    \right)^{-1/2},
\label{u5}
\end{equation}

\noindent which corresponds to the expressions calculated by \citet{ulrich76}.

\section{Streamlines and density profiles}
\label{streamlines}

\begin{figure}
  \includegraphics[scale=0.48]{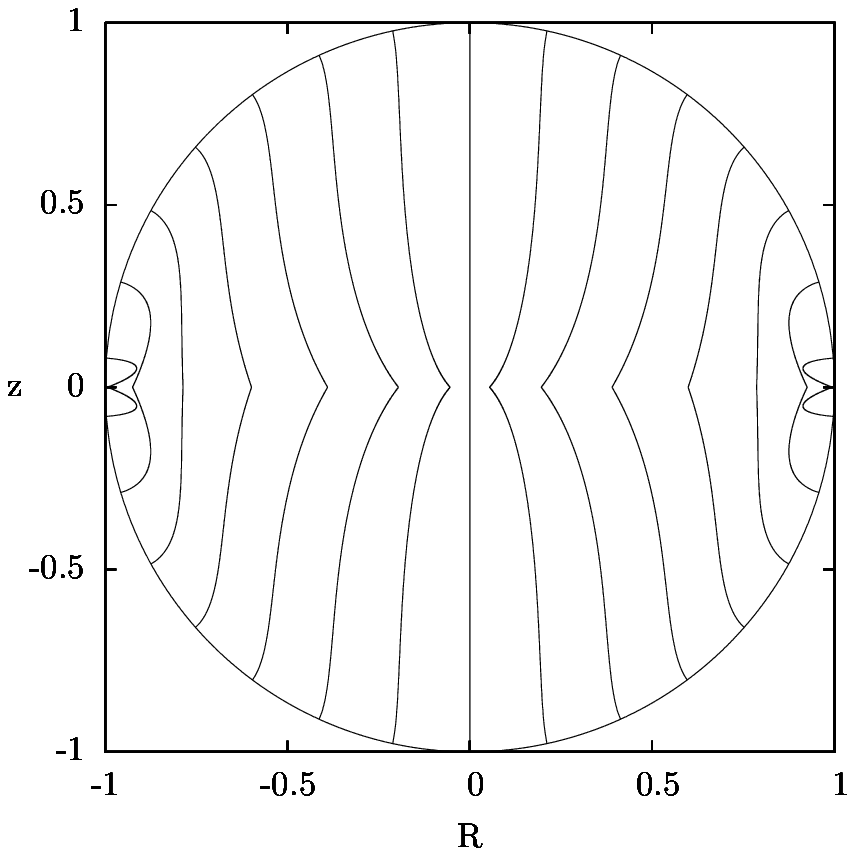}
  \includegraphics[scale=0.48]{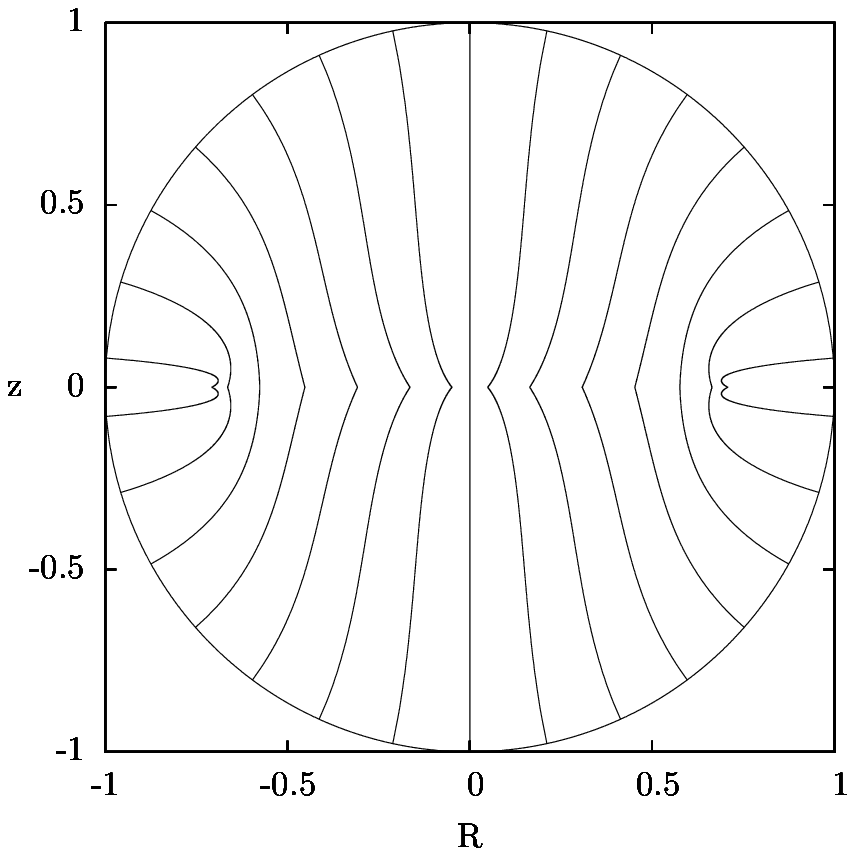} \\
  \includegraphics[scale=0.48]{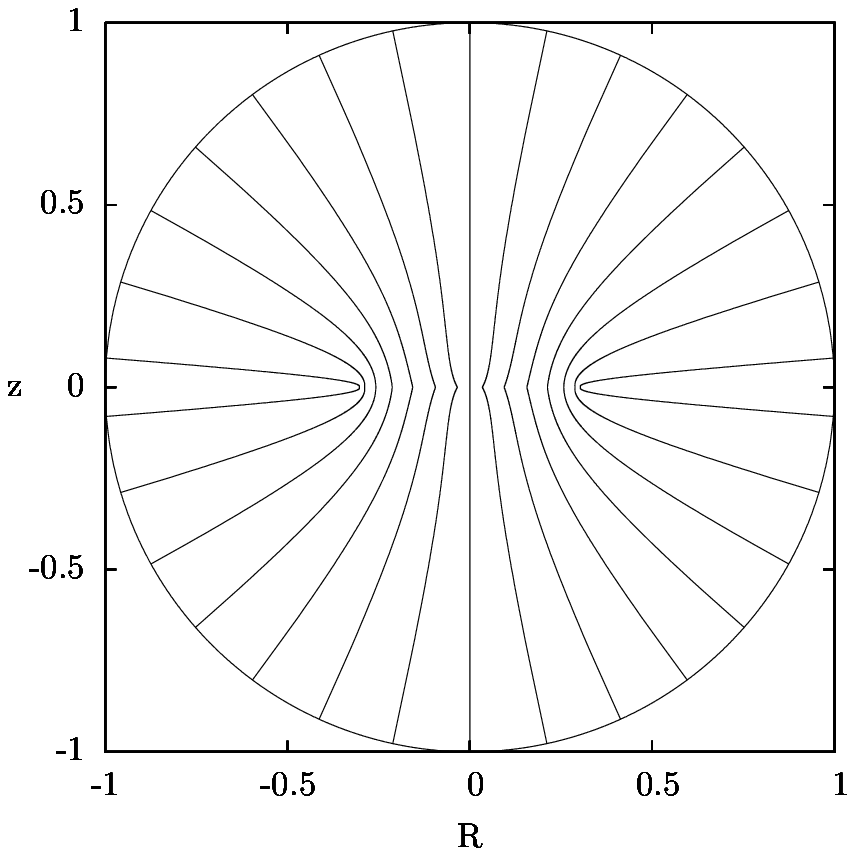}
  \includegraphics[scale=0.48]{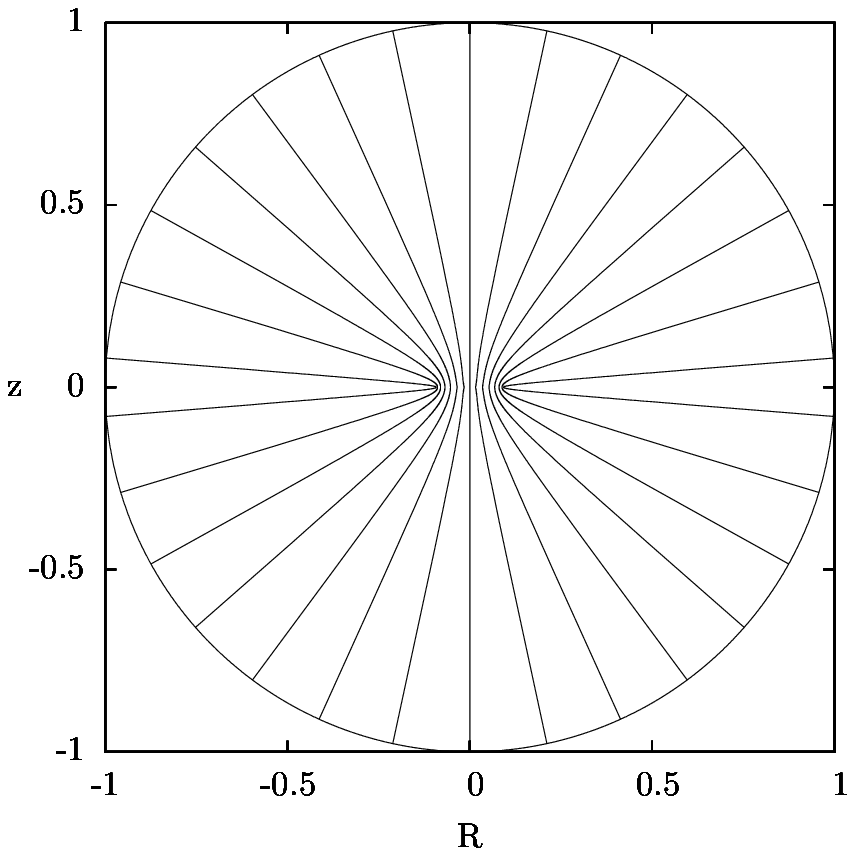}
  \caption[Streamlines of the flow]{The figure shows projected streamlines 
  of the flow for different values of the parameter \( \nu \)
  (see text) at an arbitrary azimuthal angle \( \phi = \text{const} \)
  in such a way that the parameter \( \mu = 1.28 \).  From
  left to right, top to bottom, these values are 
  \( \nu = 0.28,\ 0.8,\ 3.2,\ 12.8\).  These projections
  do not correspond to the real streamlines of the flow (see
  section~\ref{streamlines}).  
  Distances are measured in units of the cloud's border \( r = r_0 = 1/\mu \).}
\label{fig01}
\end{figure}

  Figure~\ref{fig01} shows projected streamlines for different values
of the parameters \( \mu \) and \( \nu \) with height \( z \) and radial
coordinate \( R:= r \sin \theta \).  Each dot on a fixed curve correspond
to the intersection of the real streamline and the plane \( \phi =
\text{const} \) as it is being swept from \( \phi = 0 \) to \( \phi =
\pi / 2  \). In other words, each line is the intersection of a family
of real streamlines all having the same \( \theta_0 \), with a fixed
plane \( \phi = \text{const} \).  The reason of doing this is because
direct projections of streamlines on a plane \( \phi = \text{const}
\) show false intersections of the streamlines on the obtained plots.
Figure~\ref{fig01} shows the projected streamlines for a fixed \( \mu \)
and  different values of \( \nu \).

\begin{figure}
  \includegraphics[scale=0.48]{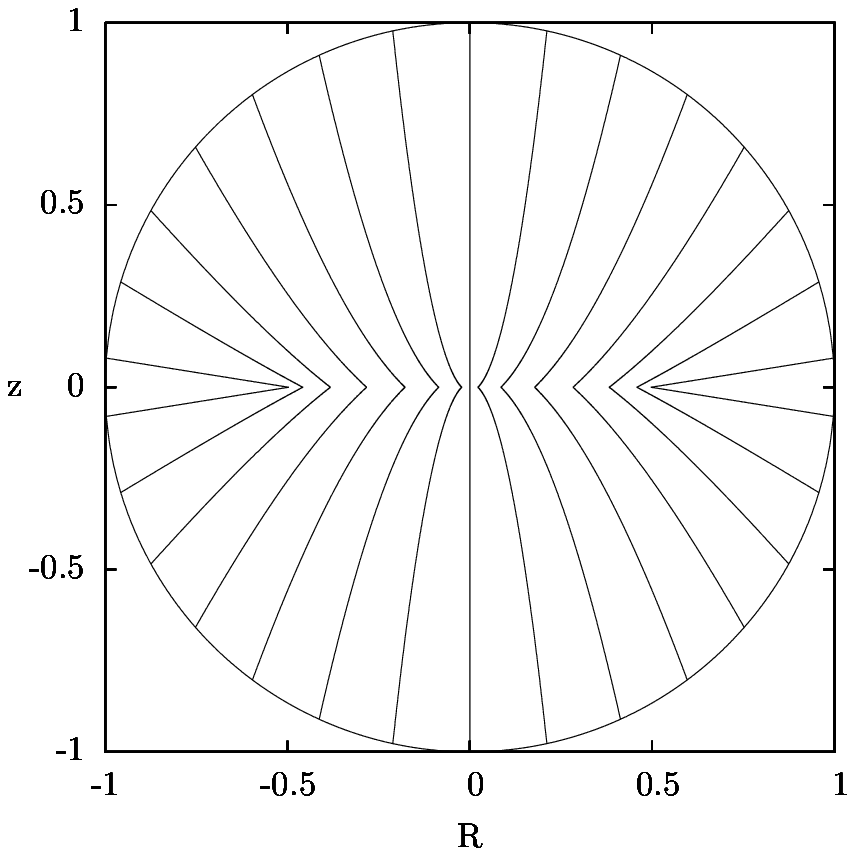}
  \includegraphics[scale=0.48]{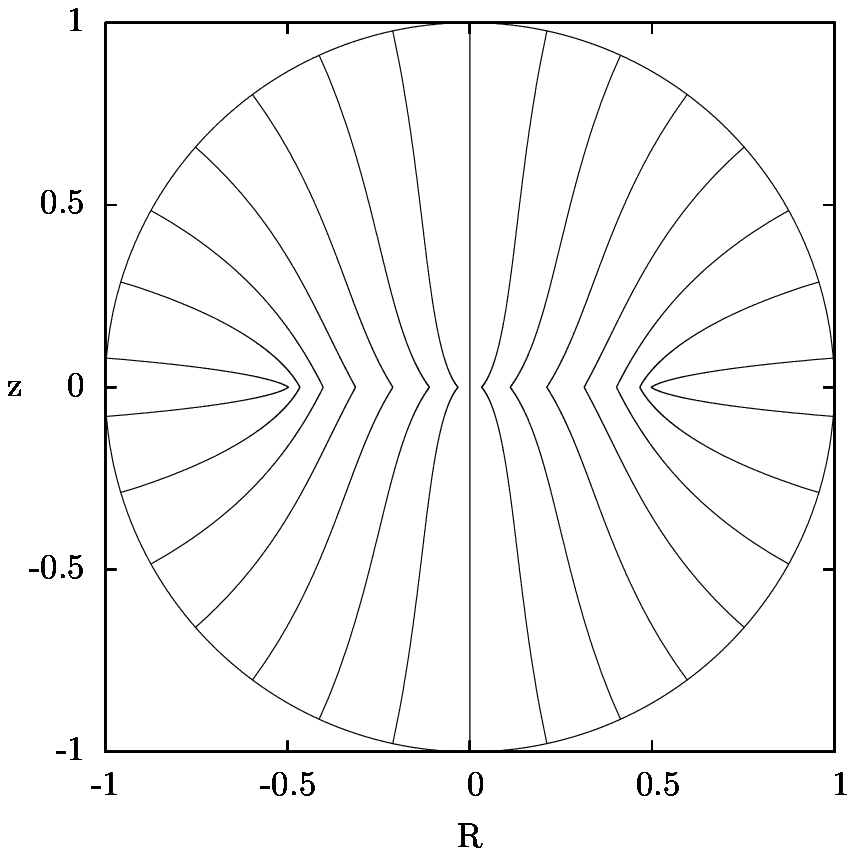} \\
  \includegraphics[scale=0.48]{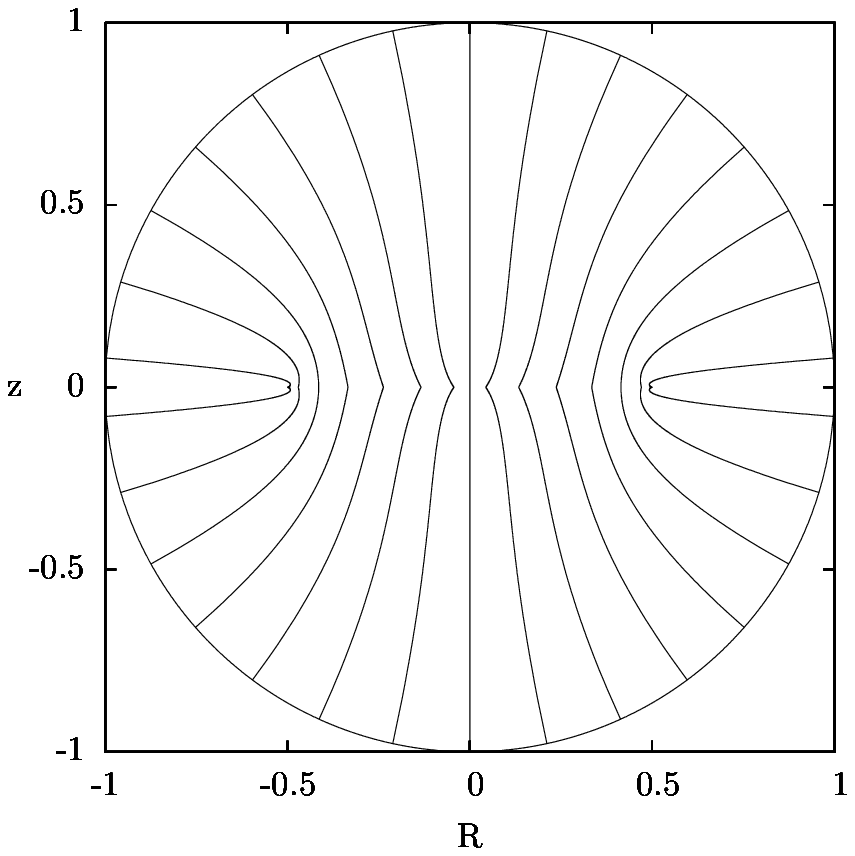}
  \includegraphics[scale=0.48]{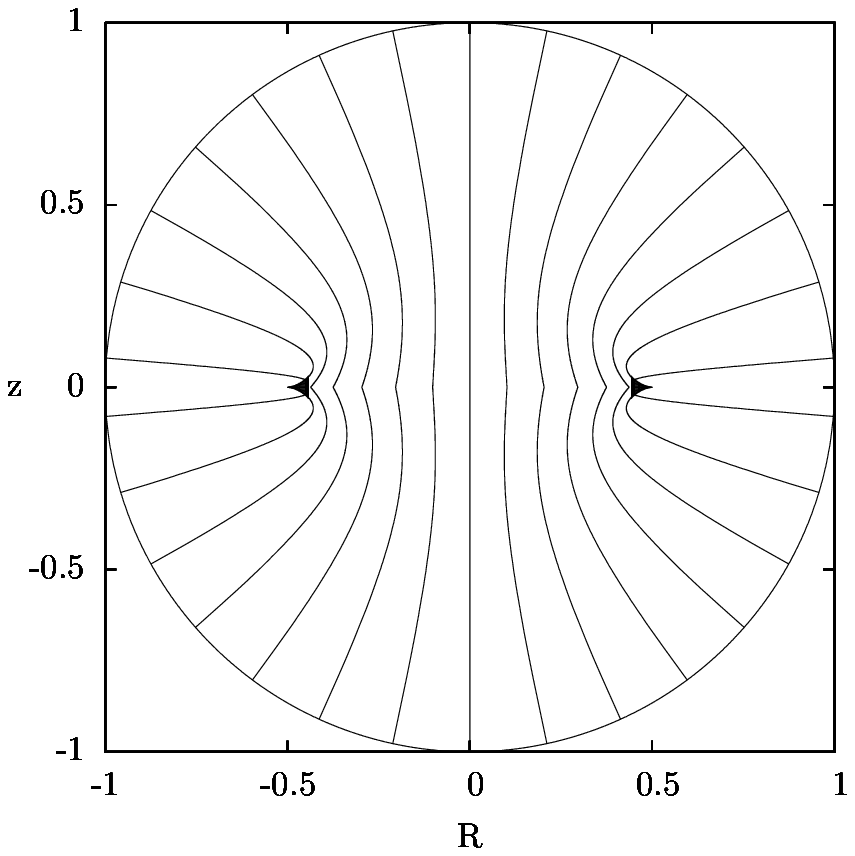}

  \caption[Streamlines of the flow]{The figure shows projected streamlines 
  of the flow for different values of the parameters \( \mu \) and \( \nu \)
  (see text) at an arbitrary azimuthal angle \( \phi = \text{const} \)
  in such a way that the radius of the disc has a value
  \( r_\text{d} = 1/2\mu \). From
  left to right, top to bottom, these values are \( \mu = 0.5,\
  0.81,\ 1.31,\ 51.01\), \( \nu = 0,\ 0.64,\ 1.62,\ 101\).  These projections
  do not correspond to the real streamlines of the flow (see
  section~\ref{streamlines}).  The filled ``triangles'' drawn in the bottom
  plots are zones out-of-range where streamlines of the flow intersect.  This
  occurs because the flow has been assumed to be ballistic.
  Distances are measured in units of the cloud's radius \( r = r_0 = 1/\mu \).}
\label{fig02}
\end{figure}

  Figure~\ref{fig02} shows also projected streamlines of the flow for a
disc's radius of one half of the original cloud's radius 
(\( r_\text{d} = 1/2\mu \)).
The bottom panels in Figure~\ref{fig02} have ``triangles'' drawn on them.
These are zones out-of-range in the model due to real intersections of the
accretion streamlines.  Note that this behaviour appears because the 
model is ballistic.  We have explored numerically in which
cases these intersections occur and it turns out that for \( r_\text{d}
= 1/2\mu \), the streamlines show intersections for \( \nu \gtrsim  1 \).
From values of \( \nu \gtrsim 50 \) intersections appear but the out-of-range 
region is kept fixed. 
  
\begin{figure}
  \includegraphics[scale=0.46]{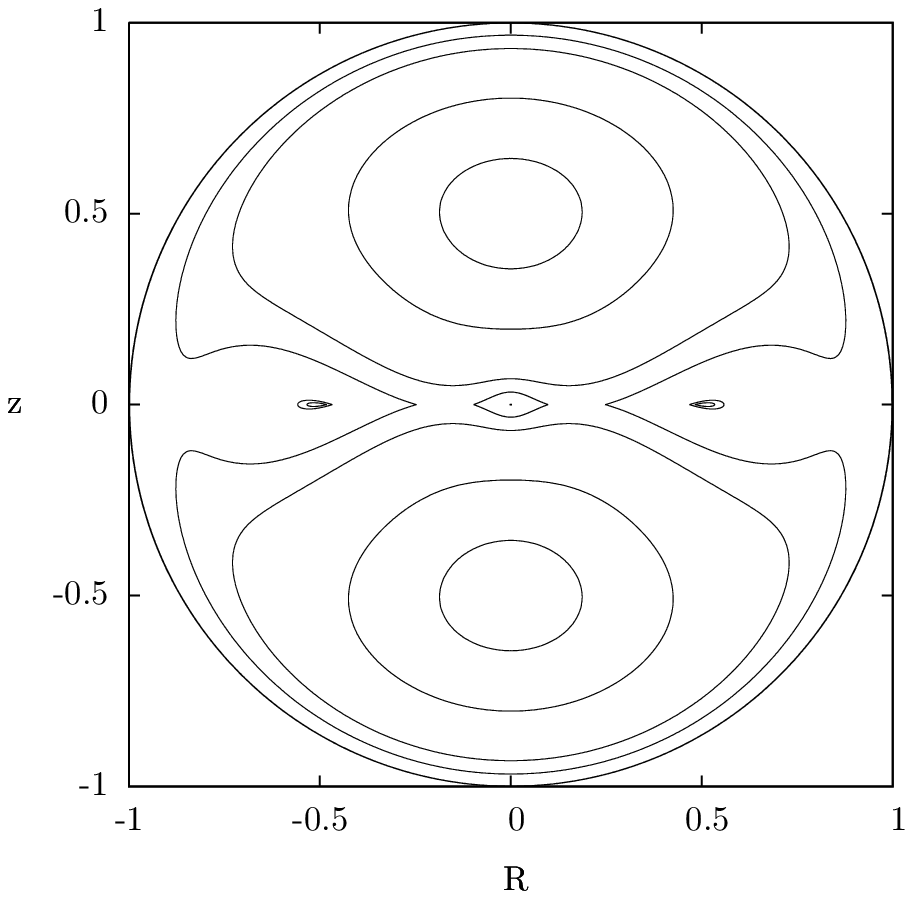}
  \includegraphics[scale=0.46]{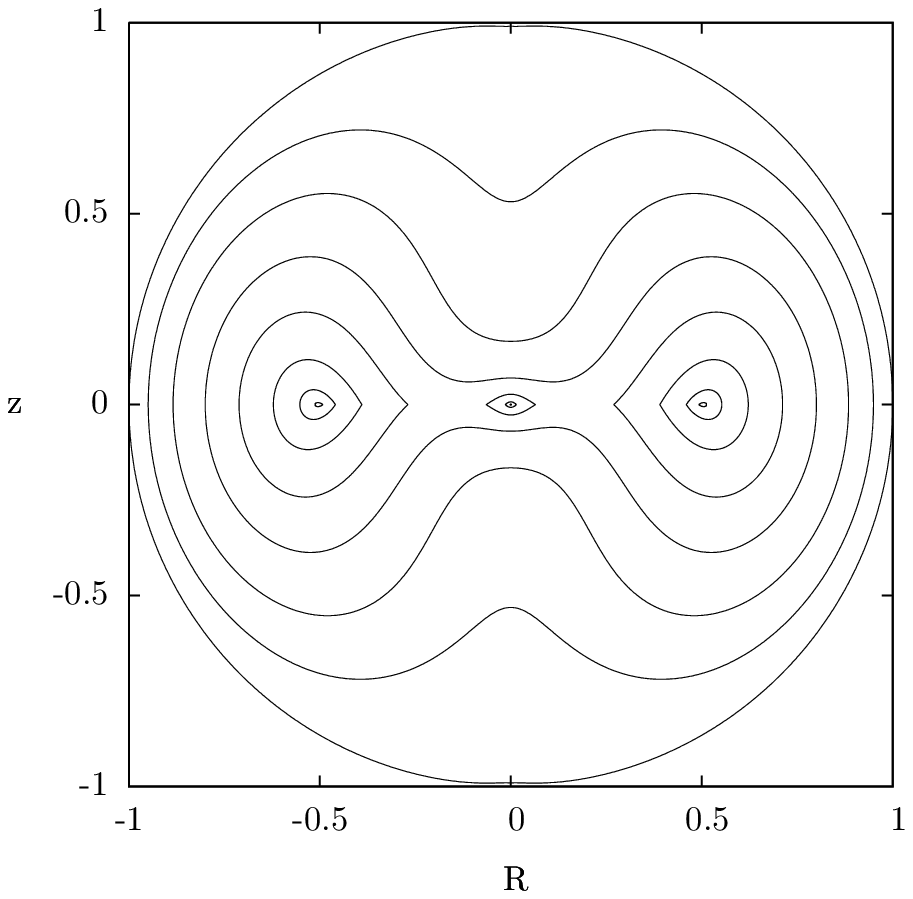} \\
  \includegraphics[scale=0.46]{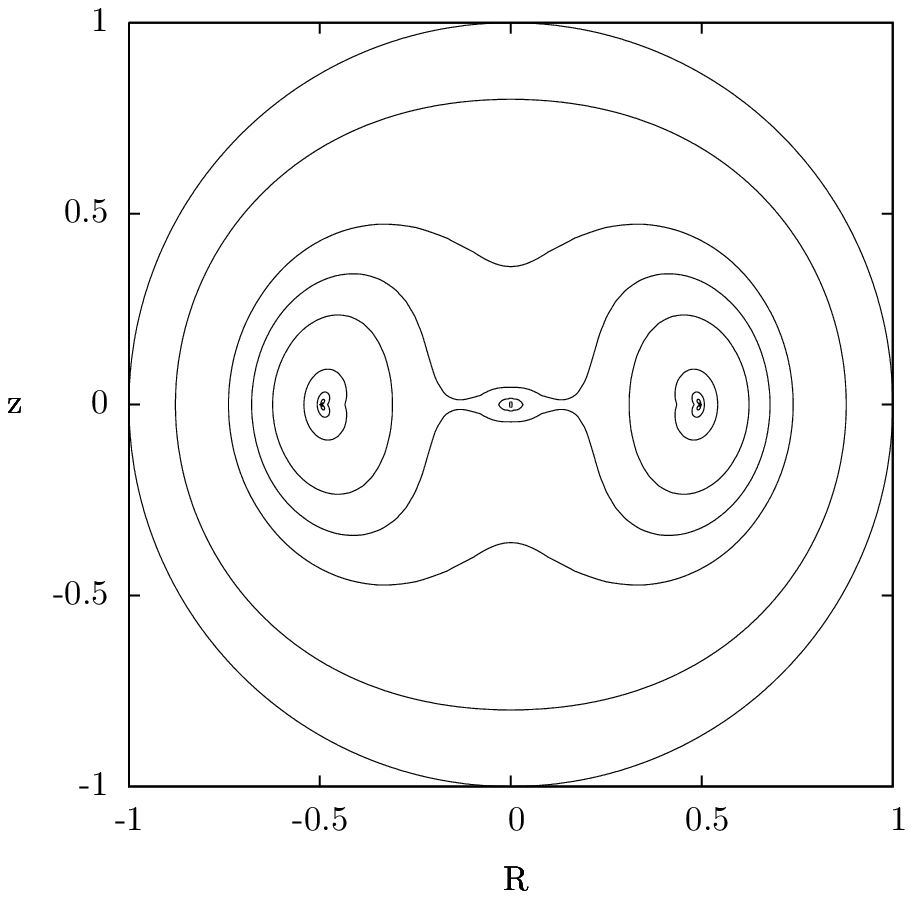}
  \includegraphics[scale=0.46]{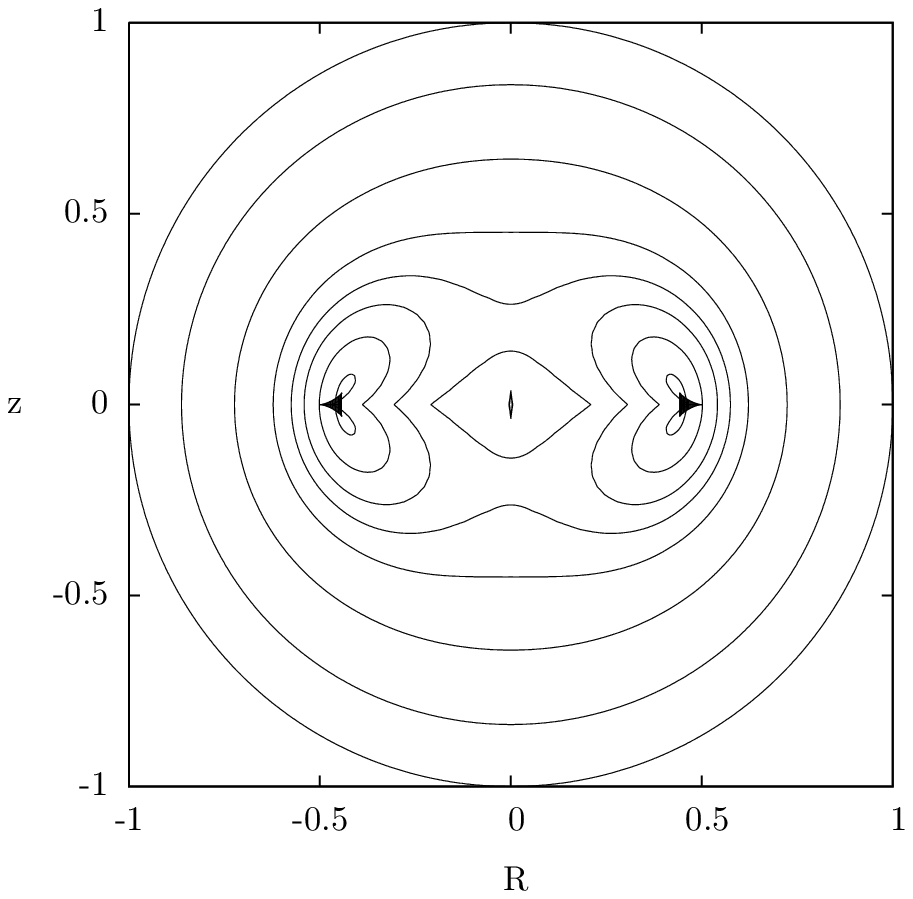}
  \caption[Density isocontours]{The figure shows 
  logarithmic particle number density \( n \) isocontours in 
  units of \( n_\text{u} \). From left to right, top to
  bottom the values of the parameters of the flow are the same as in
  Figure~\ref{fig02} and the values of the isocontours for each panel 
  are given by \( \{ 0.32,\ 0.4,\ 0.6,\ 0.75,\ 1.5,\ 1.75 \} \), 
  \( \{ 0.3,\ 0.34,\ 0.4,\ 0.5,\ 0.65,\ 0.9,\ 1.3,\ 2.0 \} \), 
  \( \{ 0.0,\ 0.13,\ 0.33,\ 0.45,\ 0.6,\ 1.0,\ 1.5,\ 2.0 \} \) and \( \{ 
   -1.0,\ -0.85,\ -0.65,\ -0.45,\ -0.32,\ -0.2,\ 0.0,\ 0.5 \} \) 
  respectively.  The dark regions on the bottom panels are the
  out-of-range zones.  
  Distances are measured in units of the cloud's radius
  \( r = r_0 = 1/\mu \).}
\label{fig03}
\end{figure}

\begin{figure}
  \includegraphics[scale=0.68]{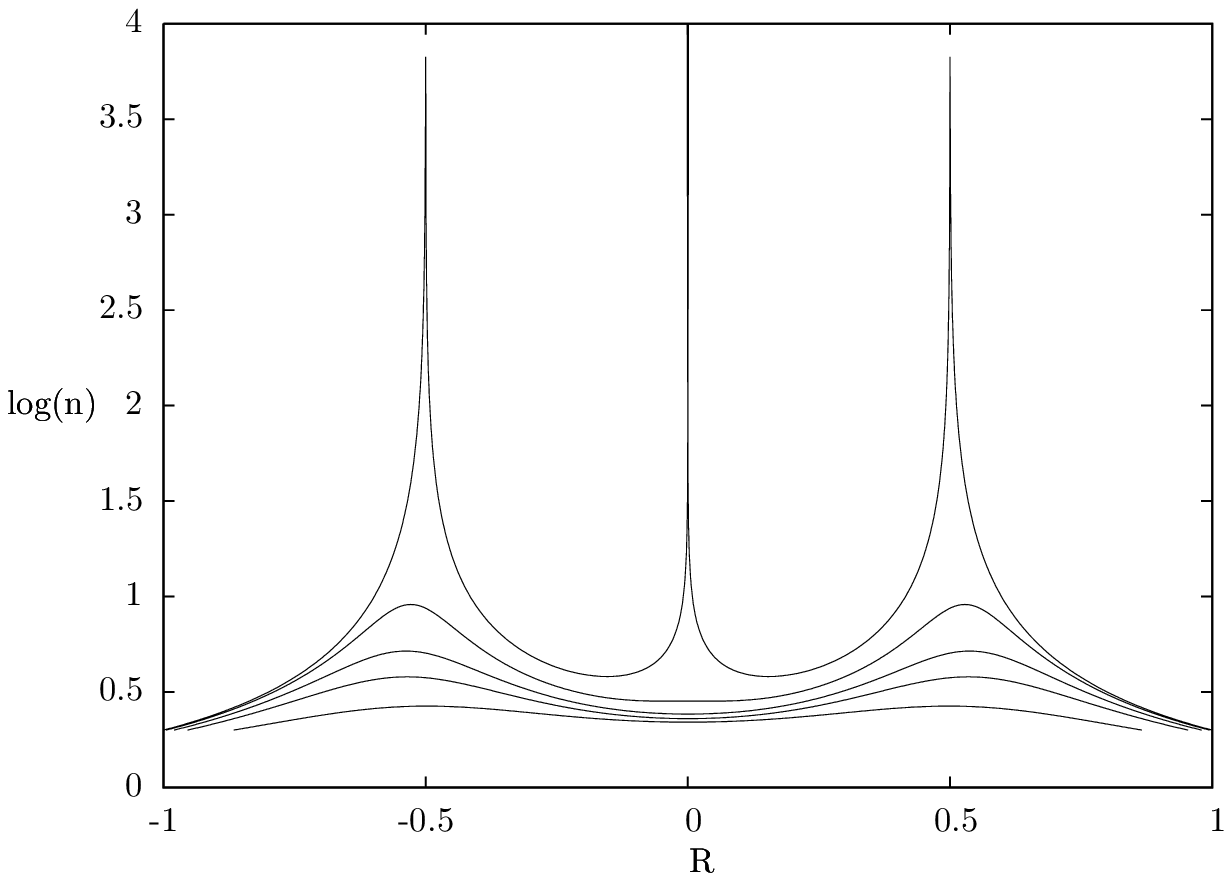}
  \includegraphics[scale=0.68]{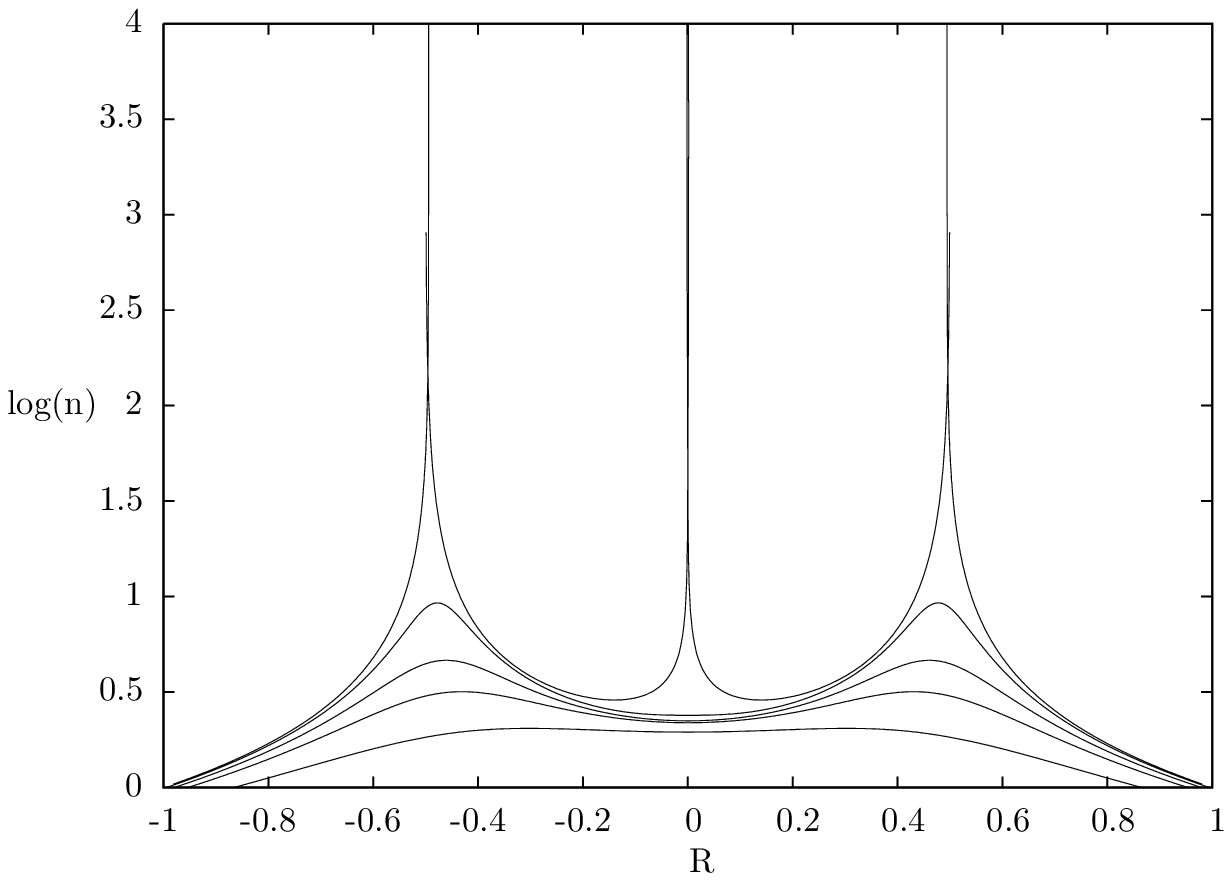} 
  \caption[Density of the flow]{Both panels show density profiles for a
  fixed azimuthal angle, i.e. \( \phi = \text{const} \) and a fixed height
  \( z = 0,\ 0.1,\ 0.2,\ 0.3,\ 0.5 \) measured from the equatorial plane.
  The figures were drawn using values of \( \nu=0.64,\ 1.62 \) respectively
  for a fixed value of the radius of the disc \( r_\text{d} = 1/2\mu \). Note
  that for a height \( z= 0 \) the particle number density diverges at \( r
  = 0, \ \pm 0.5 \). Distances are measured in units of the cloud's radius
  \( r = 1/\mu \).}
\label{fig04}
\end{figure}

  Figures~\ref{fig03} and~\ref{fig04} show density isocontours and
density profiles for a fixed height measured from the equatorial plane
respectively.  The out-of-range region appears only in the bottom panels
of  Figure~\ref{fig03}, although on the left-bottom panel it appears as
a tiny region.

\begin{figure}
  \includegraphics[scale=0.71]{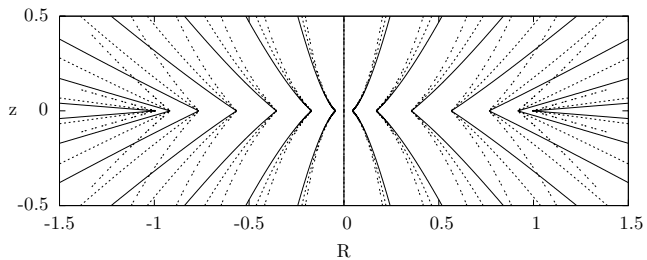}
  \caption{The plot shows a comparison between the streamlines for three
  different values of the initial cloud's radius in the case \( \nu =
  0\). These values are \( r_0/r_\text{u} = \infty,\ 2.46,\ 1.38\) for
  the cases of continuous, dotted and dotted-blank lines respectively.
  Distances are measured in units of \(r_\text{d}\).  Bear in mind that
  for different values of \( r_0 \) the disc radius \( r_\text{d} \)
  varies and so, if one is to keep \( r_\text{d} = 1 \) then different
  values of the mass of the central object and the angular momentum of
  the cloud need to be properly readjusted.  Distances in the plot are
  measured in units of the radius of the disc \( r_\text{d} \).  }
\label{fig04b}
\end{figure}

  Figure~\ref{fig04b} shows plots of the streamlines for the case 
\( \nu = 0 \). The streamlines on the plot have different values of 
the initial cloud's radius.
It is interesting to note that the different streamlines having the
same \( \theta_0 \) as initial condition, arrive to the same projected
position over the disc.

\begin{figure}
  \includegraphics[scale=0.71]{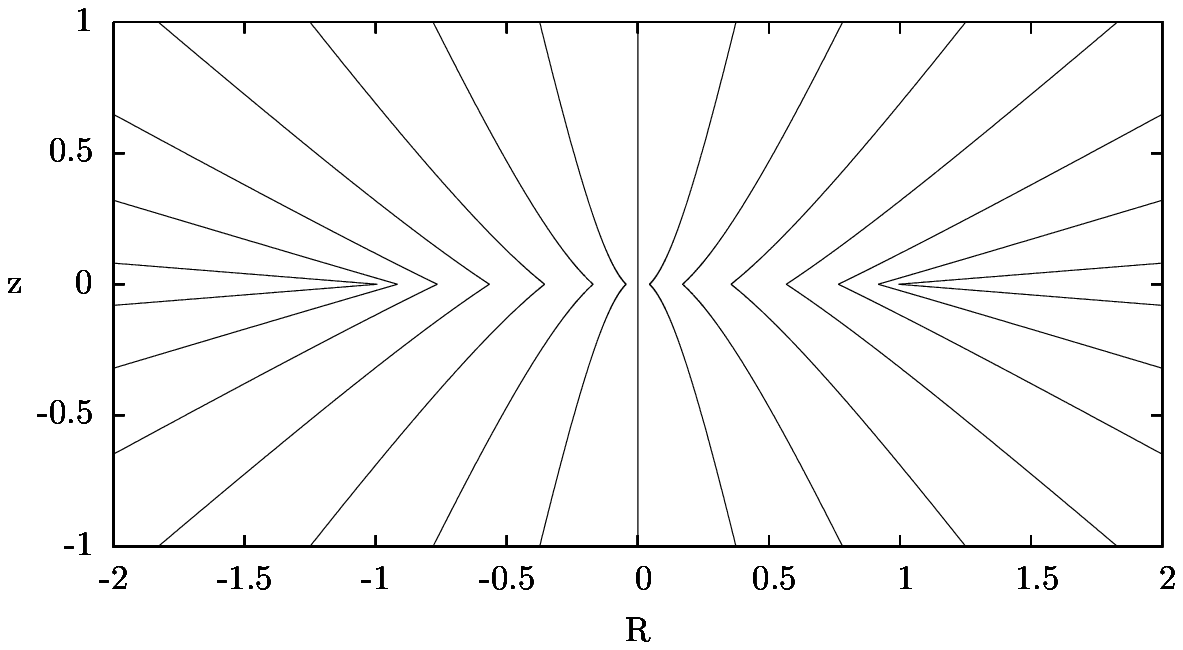}
  \includegraphics[scale=0.71]{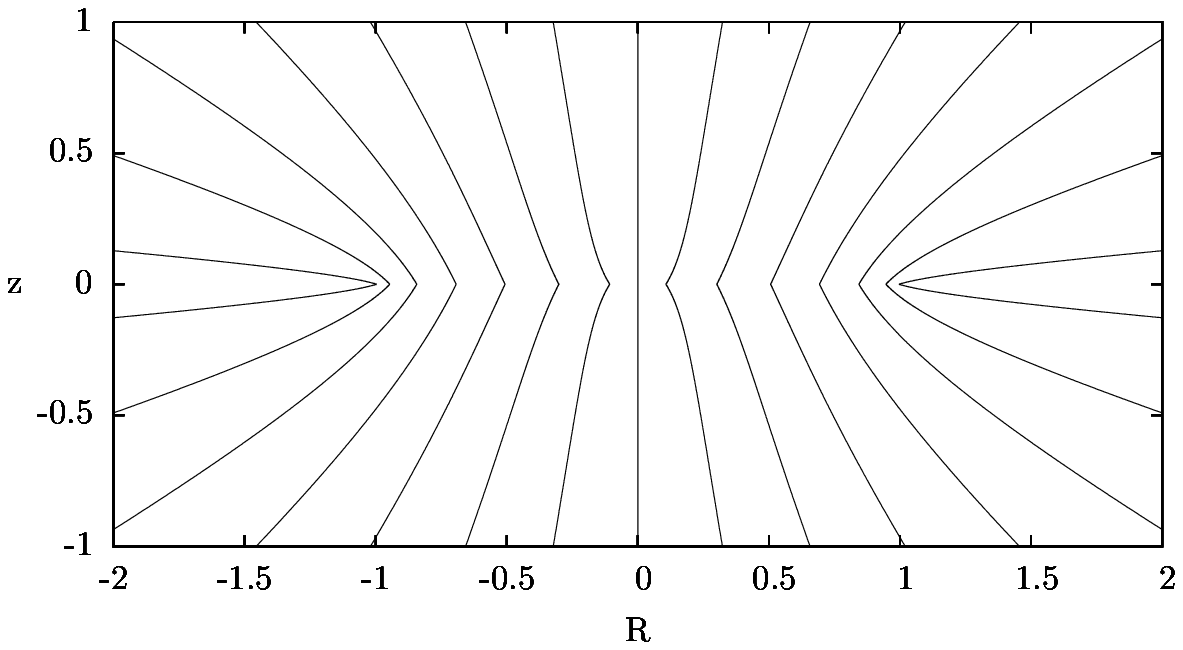} 
  \includegraphics[scale=0.71]{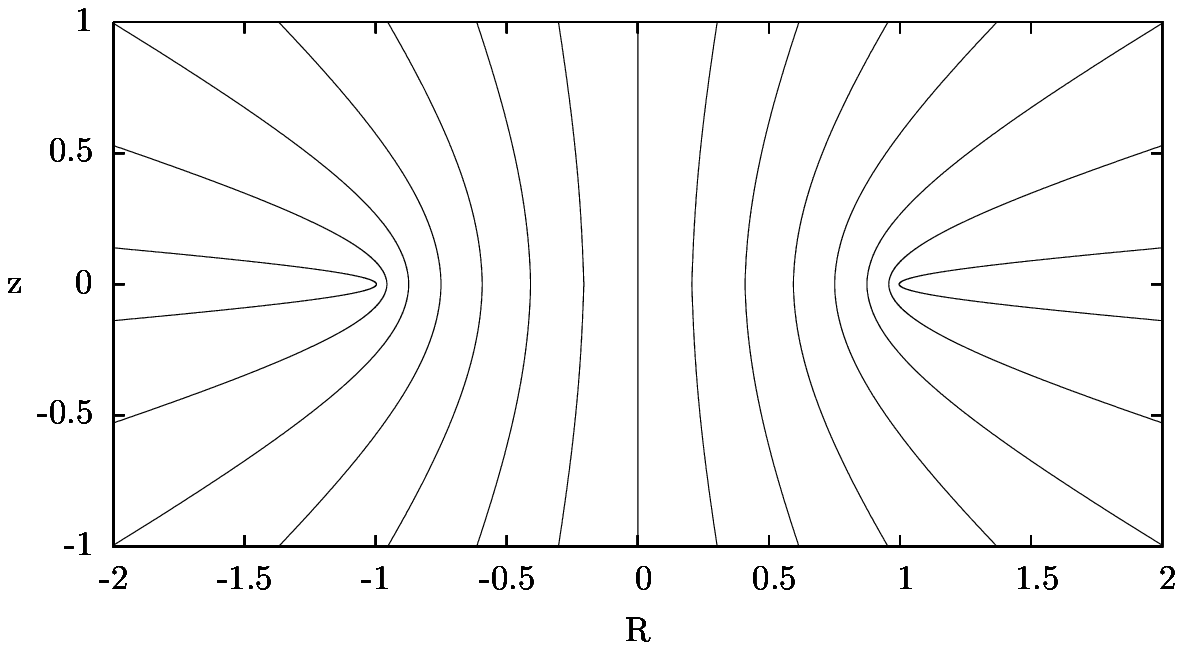}
  \caption{The plots show projections of streamlines (in the same sense as
  in Figure~\ref{fig01}) which  correspond to an infinite value of the cloud's
  radius \( r_0 \).  From top to bottom the parameter \( \nu \) has values
  of  \( 0,\ 2,\ 100\)  respectively (see Section~\ref{velocity}).  The
  top panel corresponds to \citeauthor{ulrich76}'s accretion model.
  Distances are measured in units of the radius of the disc \(
  r_\text{d}\).}
\label{fig05}
\end{figure}

\begin{figure}
  \includegraphics[scale=1.05]{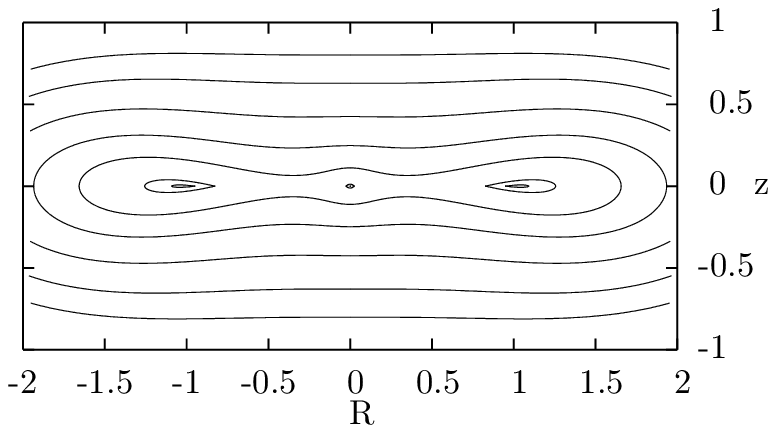} 
  \includegraphics[scale=1.05]{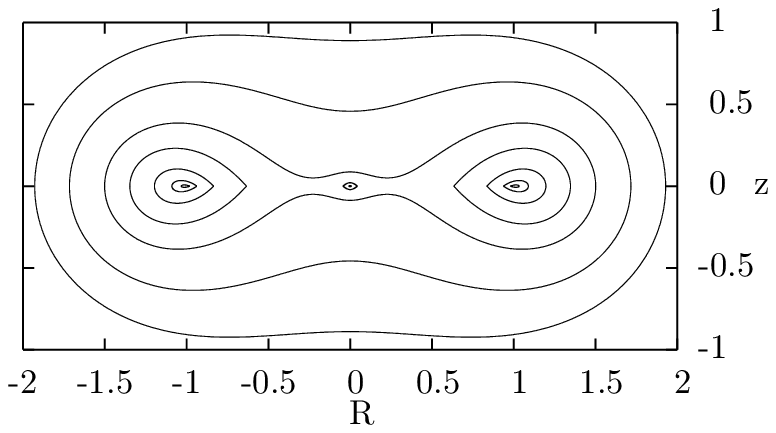}
  \includegraphics[scale=1.05]{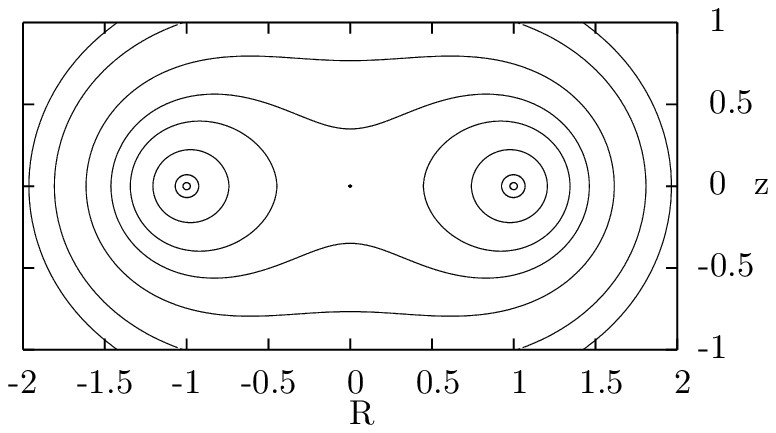} 
  \caption{The panels show the logarithmic particle number density
  isocontours measured in units of \( n_\text{u} \) corresponding
  to the same set of parameters chosen in Figure~\ref{fig05}. Distances
  are measured in units of the radius of the disc \( r_\text{d} \) and
  from top to bottom the sets of contours take the following values: \(
  \{ -0.55,\ -0.47,\ -0.35,\ -0.2,\ 0.0,\ 0.5,\ 1.0 \} \), \( \{ 0.25,\
  0.4,\ 0.6,\ 0.8,\ 1.1,\ 1.5,\ 2 \}  \),  and \( \{ 1.7,\ 1.8,\ 1.95,\
  2.1,\ 2.25,\ 2.5,\ 3,\ 3.5 \} \).   }
\label{fig06}
\end{figure}

\begin{figure}
  \includegraphics[scale=0.67]{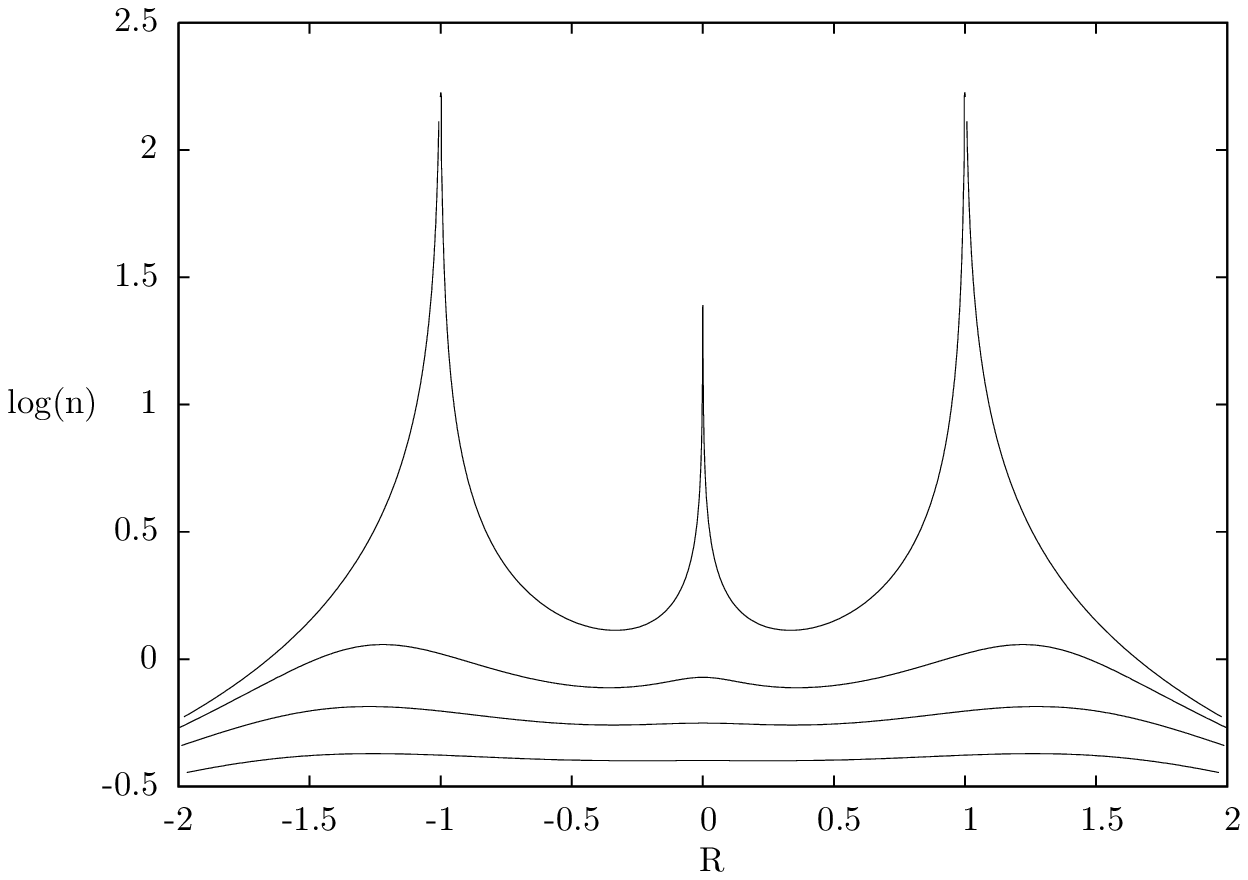} 
  \includegraphics[scale=0.67]{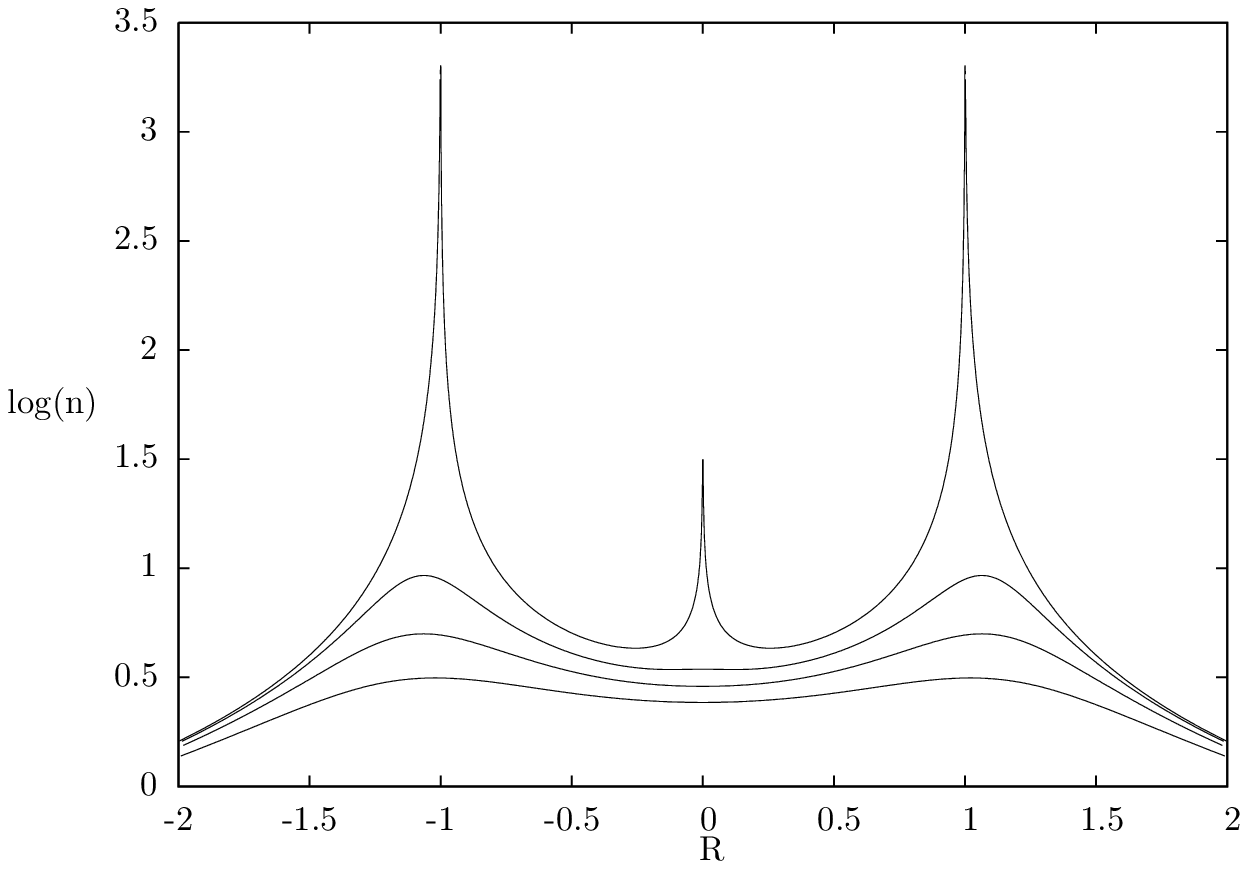} 
  \includegraphics[scale=0.67]{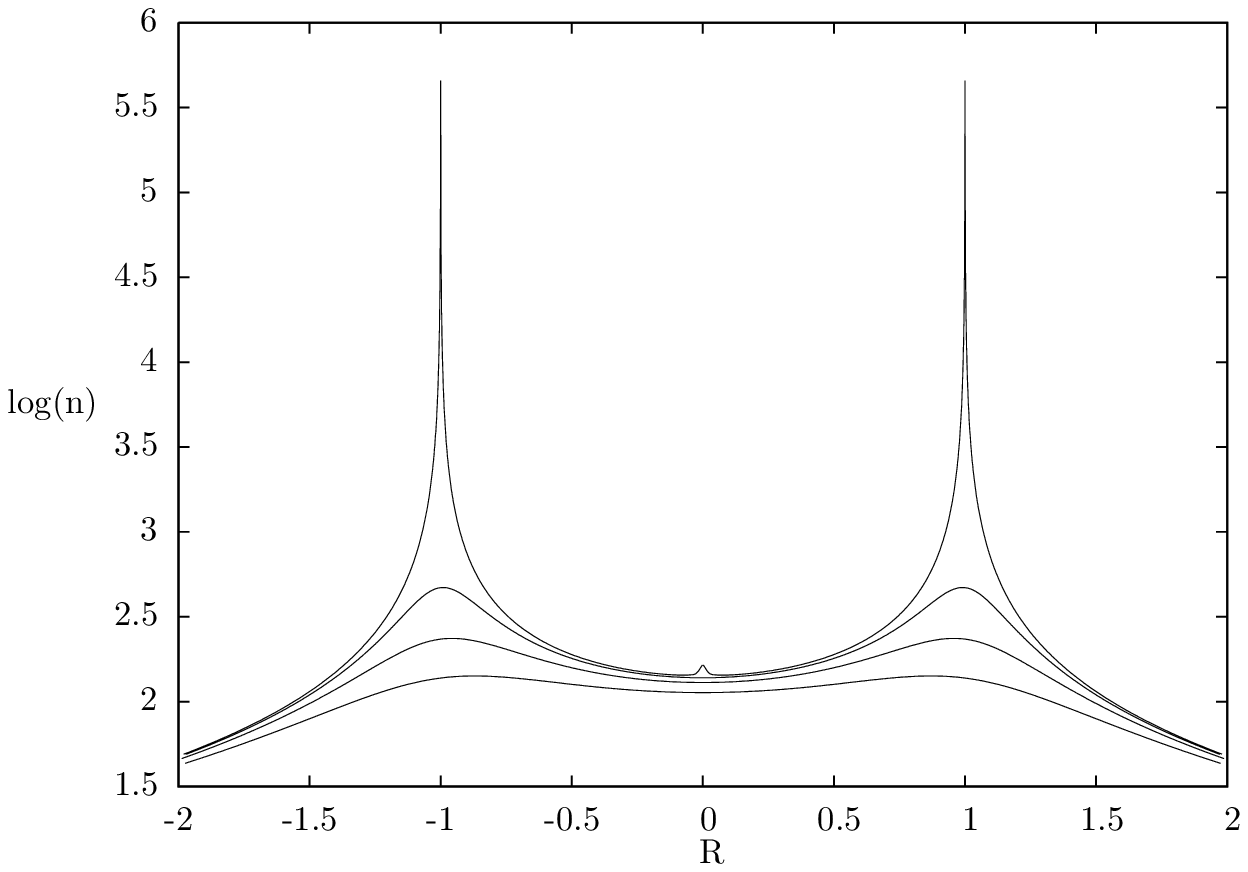} 
  \caption{From top to bottom, the figures show the logarithm of the
  particle number density of the flow in units of \( n_\text{u} \),
  corresponding to the same values of \( \nu \)  that Figure~\ref{fig05}
  has.  All plots have density as a function of the equatorial radial
  coordinate \( R \) for different fixed heights \( z =
  0,\ 0.15,\ 0.3,\ 0.5 \).  In all figures, the plot at \( z = 0 \) 
  has density divergences at \( R = 0, \pm 1 \). }
\label{fig07}
\end{figure}

  Figures~\ref{fig05}-\ref{fig07} are equivalent plots as those of
Figures~\ref{fig02}-\ref{fig04} for the case \(\mu = 0\) (i.e. \( r_\text{0}
\rightarrow \infty \)), which can be thought of as the first natural
modification to \citeauthor{ulrich76}'s model, all of them depending
on the single dimensionless parameter \( \nu \).  Note that in all these
cases, there is no intersection of the flow streamlines. This appears to
indicate that as long as the cloud reaches a sufficiently large radius
\( r_0 \), with respect to the accretion disc radius \( r_\text{d} \),
the intersections disappear.  Indeed, we have tested different accretion
models varying \( r_\text{d}/r_0 \) from \( \sim 0.0 \) to \( 1.0 \)
and the previous statement turns out to be correct.  As an example,
a value of \( r_\text{d}/r_0 \approx 0.08  \) represents the threshold
of intersections of streamlines for values of \( \nu = 10.0 \).

  From Figures~\ref{fig05} and~\ref{fig06} it is clear that the particle
number density does not accumulate that much in the origin as \( \nu \)
increases.

\section{Astrophysical applications}
\label{application}

  From the point of view of Spectral Energy Distribution (SED)
modelling of protostellar cores, one of the main ingredients required
for a radiative transfer calculation is a specification of the density
profile in the core. Many authors use Ulrich's model to achieve this
\citep[see for e.g.][]{kenyon93,butner91,adams86,whitney03}. A radiation
transfer simulation requires a lot of information: luminosity of the
star, composition, abundance and optical properties of the material
in the core, in terms of its location with respect to the star. Thus,
a full analysis of the SED produced by the use of the density given in
this article is beyond its scope, but must be consider in the future.

  In order to give a qualitative idea of the difference in the
results, using either Ulrich's density or the density calculated in
Section~\ref{particle-number} of this article, we note that less density
means more luminosity arriving from the object.  The main argument is
that the optical depth, which is proportional to the surface density calculated
including all the material along a line of sight, is an element that
give us an idea of the SED produced by the core.

\begin{figure}
  \includegraphics[scale=0.67]{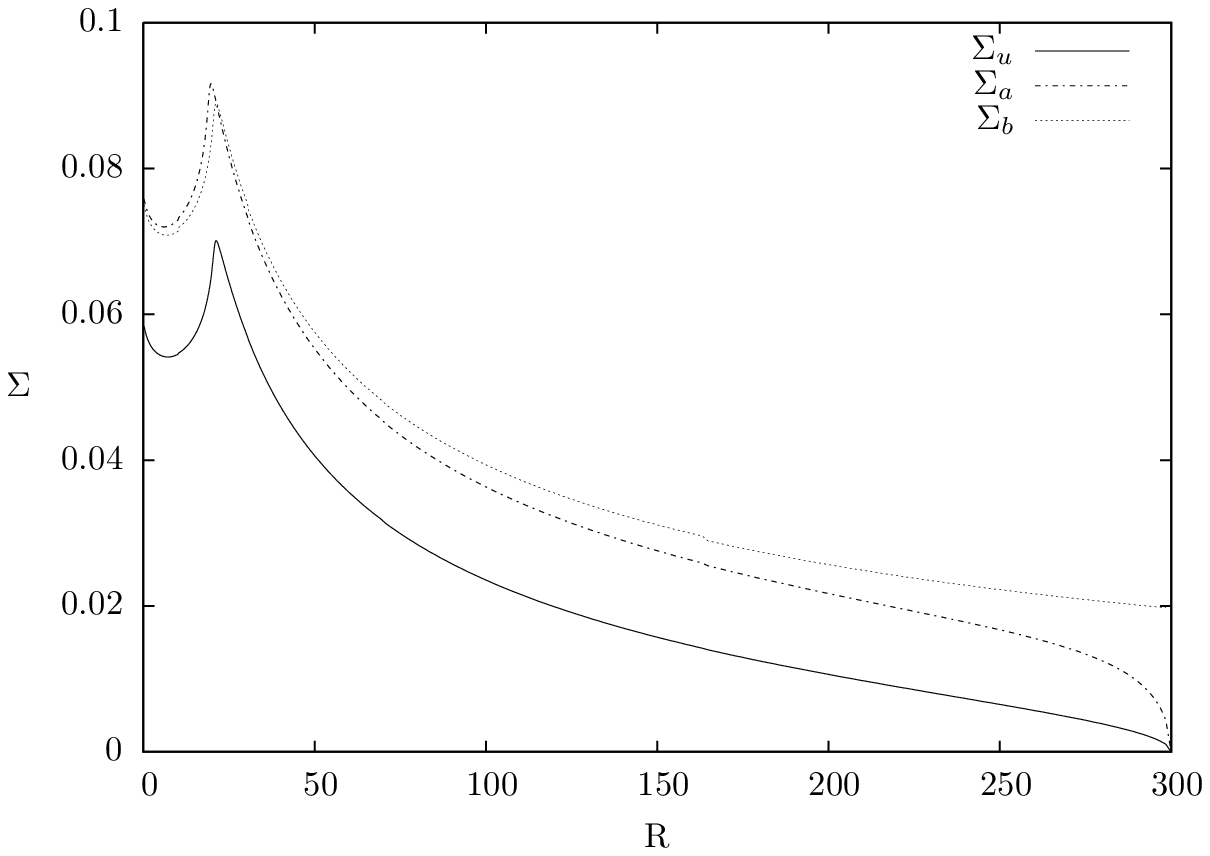}
  \includegraphics[scale=0.67]{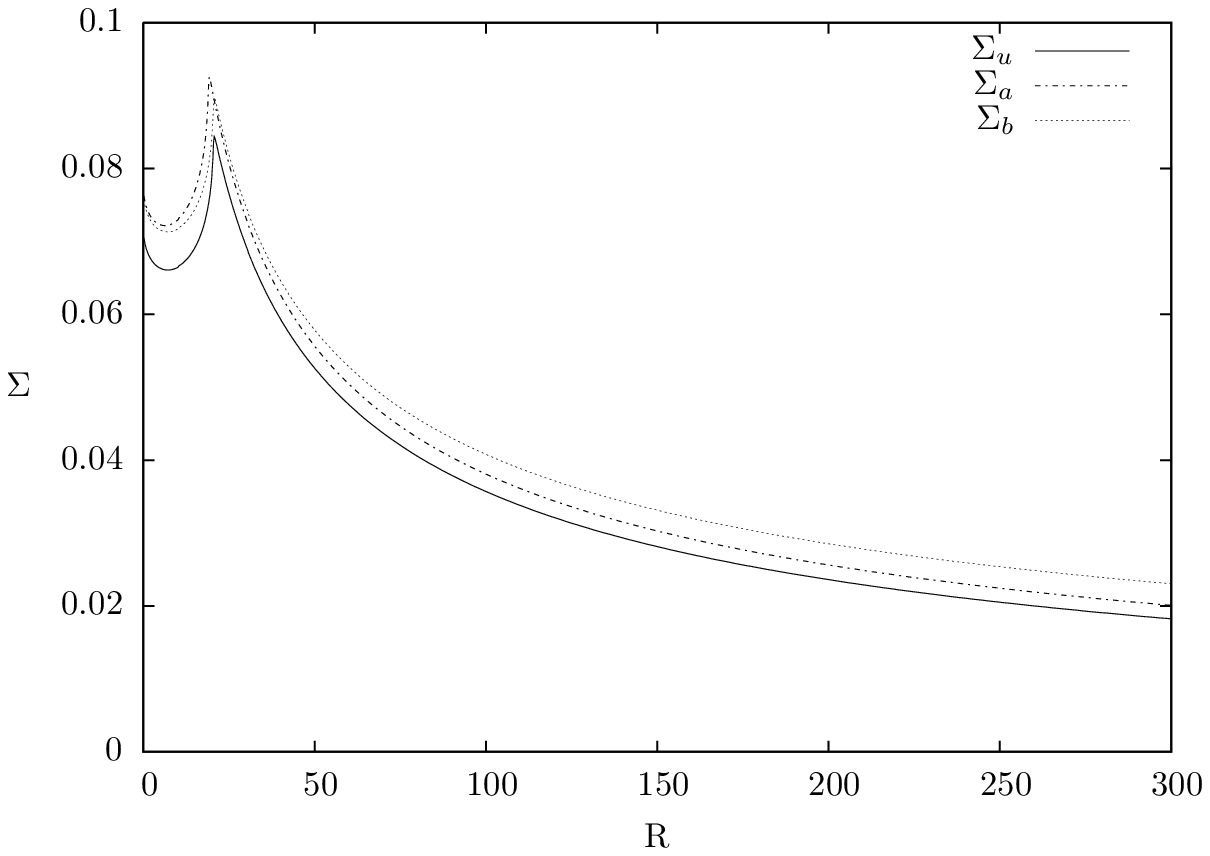}
  \includegraphics[scale=0.67]{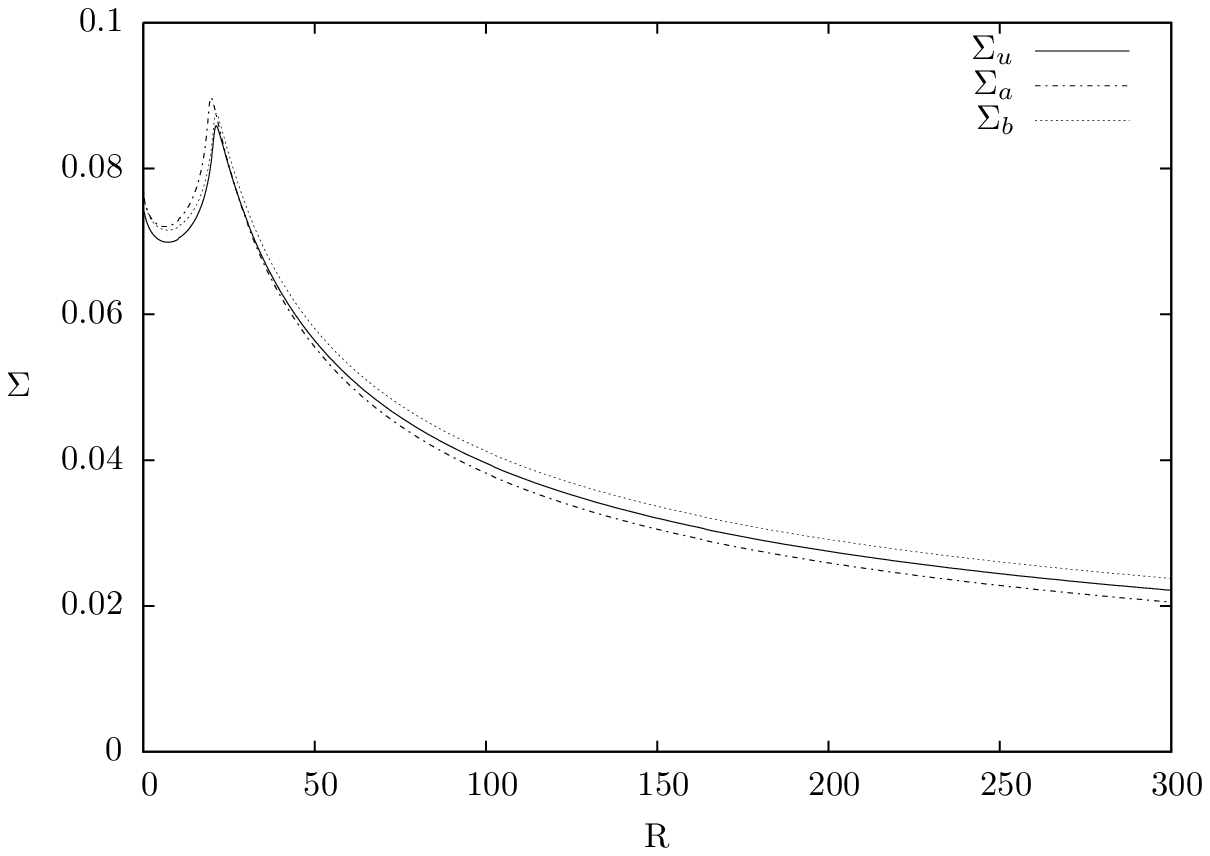}   
  \caption{The figure shows axisymmetric surface density \( \Sigma \)
	   plots as a function of the radial coordinate \( R \) for
	   a disc viewed pole-on, using an Ulrich's model for which
	   the surface density is given by  \( \Sigma_\text{u} \)
	   and from the results of this article with \( \Sigma_a \)
	   for cases with \( v_{r_0} = 0 \) and \( \Sigma_b \) for
	   \( v_{r_0} = 5 \times 10^{4}  \, \text{cm}/\text{s}  \).
	   From top to bottom, we have chosen three examples in such a
	   way that $r_\text{u} = 20 \, \text{AU} $ and the corresponding
	   sets of boundary conditions are given by:  (a) $r_{0}=300 \,
	   \text{AU}$, $\Omega=10^{-11} \, \text{s}^{-1}$, (b) $r_{0}=3000
	   \, \text{AU} $, $\Omega=10^{-13} \, \text{s}^{-1}$, and (c)
	   $r_{0}=30000 \, \text{AU}$, $\Omega=10^{-15} \, \text{s}^{-1}$.
	   The other common parameters on all models are given by $M=1 \,
	   \text{M}_{\odot}$ and  $\dot{M}=10^{-6} \, \text{M}_{\odot}
	   \, \text{yr}^{-1}$.	Distances on the plot are measured in
	   astronomical units and the surface density in \( \text{g} /
	   \text{cm}^{-2} \). }
\label{fig15}
\end{figure}

  For a disc viewed pole-on, in Figure~\ref{fig15} we present the
axisymmetric surface density \( \Sigma \) as a function of the radial
coordinate, for Ulrich's model (\( \Sigma_\text{u} \)) and from this paper
(\( \Sigma_a \) for \( v_{r_0} = 0 \) and \( \Sigma_b \) for \( v_{r_0}
= 5 \times 10^{4} \, \text{cm}/\text{s} \), according to the results of
\citet{hennebelle04}) .  This comparison is done for 3 representative
cases of $r_{0}$ and $\Omega=h_0/r_0$, such that $r_\text{u} = 20
\, \text{AU} $ is the same for all. We take three different sets of
boundary conditions: (a) $r_{0}=300 \, \text{AU}$, $\Omega=10^{-11}
\, \text{s}^{-1}$, (b) $r_{0}=3000 \, \text{AU} $, $\Omega=10^{-13} \,
\text{s}^{-1}$, and (c) $r_{0}=30000 \, \text{AU}$, $\Omega=10^{-15} \,
\text{s}^{-1}$.  Note that in all three cases, \( h_0 \) is kept fixed
since for all models the radius \( r_\text{u} \) has been chosen to have
a fixed value.  The other common parameters on these models are given
by the mass of the central star $M=1 \, \text{M}_{\odot}$ and the mass
accretion rate $\dot{M}=10^{-6} \, \text{M}_{\odot} \, \text{yr}^{-1}$.
The most typical astronomical case is (b) in accordance with the
observations of \citet{benson89,jijina99}.

  In Figure~\ref{fig15}, one can see that Ulrich's surface density
$\Sigma_u$ is smaller than the one calculated using our model,
in the inner regions of the core (few times $r_\text{d}$). The difference
between both curves increases as $r_{0}$ decreases. Thus, the flux of
radiation in the typical frequencies produced in this region are larger
in the former than in the latter. Of course, a full analysis of this
problem awaits an appropriate simulation, but since case (b) is quite
common in accordance to observations, we can see that the modification to
Ulrich's model is quite important.  From the figures it follows that 
even if \( v_{r_0} \) has a null value, our model shows significant
deviations as compared to the ones predicted by an Ulrich approach.

  The last part of this section is intended to show the relevance of having
flexible boundary conditions for the accretion flow studied in this
article.  The collapse of non-rotating clouds was studied by
\citet{larson69,penston69,hunter77}. As an example, \citet{whitworth01}
built a model, in order to fit lifetimes and accretion rates for
class $0$ protostars (this stage ends until half the final mass of the
central object is assembled). In this model, the density differs from
the singular isothermal sphere (SIS), cf. \citet{shu77}, because in the
former, the density has an inner flat profile. Moreover, observations of
protostellar cores \citep{ward-thompson99} suggest that initial conditions
for protostellar collapse depart from SIS. 

At this point, it is clear that the boundary conditions in a collapsing
core depend on the environment. Dynamical collapses are present in regions
like $\rho$ Ophiuchi and Perseus clusters, where the collapse and 
subsequent star formation is triggered by external agents 
\citep{motte01}. \citet{henriksen97} stress that
strong accretion observed in $\rho$ Oph is related to the impact of an
interstellar shock wave. In regions like Taurus it is possible that cores
collapse by themselves without an external perturbation and so, the mass 
accretion rate is closer to the value calculated for the SIS
\citep{shu77}.

Triggered star formation by an external mechanism is throughfully studied
by \citet{hennebelle04}. This work presents numerical simulations of
a rotating core, with high spatial resolution, for the study of the
formation of a disc. The relevance of the perturbation is measured with
their parameter $\phi$, which is the ratio between the time-scale in
which the internal pressure doubles, to the sound-crossing time of the
core. A high value of $\phi$, represents a slow compression (spontaneous
collapse); the opposite is found with a small value, i.e. fast compression
(strongly externally triggered).

The difference between both cases is clearly seen in the size-scale of the
disc. In the first one, the disc evolves to a two-armed spiral pattern
without fragmentation. Ring formation is the outcome usually found in
the second case. Most of the material reaching the centre has too much
angular momentum and so, it moves outwards and establishes a dense ring
\citep{nagel07}. This configuration promotes fragmentation.  Despite the
fact that in this case \citet{hennebelle04} found no central protostar,
there must be situations where certainly it is present. In conclusion,
either spontaneous or externally triggered collapse can be studied with
the ballistic solution presented in this paper.

 Both, \citet{hennebelle04} and our work start from a core in solid
body rotation.  The former evolves to a core of differential rotation,
where a rapid compression implies higher velocities and densities in
the outer parts of the disc.  This fast compression intuitively means
that a larger \( v_{r_0} \) is present.  Indeed, if in our model we set
a large \( v_{r_0} \), then higher velocities and densities are obtained
in the same region.  For, substituting $\theta = \pi/2$ and $\theta_{0}$
in terms of $r$ from equation~\eqref{eq09} in equation~\eqref{eq13} it
is found that for a given \( r \), the velocity \( v_r \) and particle
number density \( n \) increase with an increasing \( \nu \). This
conclusion, expected intuitively, reinforces the idea that the solution
presented in this paper can be taken for studies of disc formation with
an analytical approach.

 The robustness of this kind of solutions is that, depending on the
particular astrophysical situation, we can choose a ballistic solution
with appropriate boundary conditions. For example, \citet{throop08}
studied numerically the accretion of material into a star-disc system
moving through the gas in a cluster. An appropriate solution, should
help to study small scales, where the star's gravity represents the
main contribution.  For the case of star formation in molecular clouds,
\citet{bate05} describe the accumulation of material as ``competitive
accretion''. The matter comes from different places with different
boundary conditions.  In this case, ballistic solutions are able to give
glimpses of the physics behind the formation of star-disc systems.

\section{Conclusion}

  We have constructed an analytic accretion flow that expands
\citeauthor{ulrich76}'s~(\citeyear{ulrich76}) model in the sense that
the radius of the rigid body rotating cloud is finite and an allowance
for a radial accretion velocity at the cloud's border is assumed.
When the radius of the cloud tends to infinity and the radial velocity
of the input flow at infinity goes to zero, the model converges to that
of \citeauthor{ulrich76}.  

  We have also shown that for real astronomical systems, where typical
sizes of the clouds are \( \sim 30000 \text{ AU} \) and for which the
initial inward velocity is null, the differences with Ulrich's model is
quite noticeable.  The problem with this hypothesis is that to achieve
a fixed accretion rate, then the density at the cloud's border needs to
be infinite.   The difference between both models becomes even greater
if one assumes an initial inward velocity different from zero. Since
this assumption makes the density not to diverge at the clouds border,
it is the more reasonable model to take for a real astronomical system.

  We are developing a relativistic model with all these properties in order
to generalise the model constructed by \citet{huerta05}. These results 
will be the subject of a subsequent article.

\section{Acknowledgements}
\label{acknowledgements}
  We thank the profound comments made by Prof. E. Ley Koo during the
preparation of this article. We are very grateful to Prof. R. Ulrich
for reviewing this article.  His comments were really useful to shape
the article in its last stages. SM and ET gratefully acknowledge support
from DGAPA (IN119203-3) at Universidad Nacional Aut\'onoma de M\'exico
(UNAM).  SM acknowledges financial support granted by CONACyT~(26344).
EN thanks support from a DGAPA (UNAM) postdoctoral fellowship.

\bibliographystyle{mn2e}
\bibliography{acc}

\begin{thebibliography}{}

\bibitem[\protect\citeauthoryear{{Adams} \& {Shu}}{{Adams} \&
  {Shu}}{1986}]{adams86}
{Adams} F.~C.,  {Shu} F.~H.,  1986, \apj, 308, 836

\bibitem[\protect\citeauthoryear{{Bate} \& {Bonnell}}{{Bate} \&
  {Bonnell}}{2005}]{bate05}
{Bate} M.~R.,  {Bonnell} I.~A.,  2005, \mnras, 356, 1201

\bibitem[\protect\citeauthoryear{{Beloborodov} \& {Illarionov}}{{Beloborodov}
  \& {Illarionov}}{2001}]{beloborodov01}
{Beloborodov} A.~M.,  {Illarionov} A.~F.,  2001, \mnras, 323, 167

\bibitem[\protect\citeauthoryear{{Benson} \& {Myers}}{{Benson} \&
  {Myers}}{1989}]{benson89}
{Benson} P.~J.,  {Myers} P.~C.,  1989, \apjs, 71, 89

\bibitem[\protect\citeauthoryear{{Bertoldi}}{{Bertoldi}}{1989}]{bertoldi89}
{Bertoldi} F.,  1989, \apj, 346, 735

\bibitem[\protect\citeauthoryear{{Bondi}}{{Bondi}}{1952}]{bondi52}
{Bondi} H.,  1952, \mnras, 112, 195+

\bibitem[\protect\citeauthoryear{{Bondi}}{{Bondi}}{2005}]{bondi05}
{Bondi} H.,  2005, {Accretion}.
The Scientific Legacy of Fred Hoyle, pp 55--57

\bibitem[\protect\citeauthoryear{{Butner}, {Evans} II, {Lester}, {Levreault} \&
  {Strom}}{{Butner} et~al.}{1991}]{butner91}
{Butner} H.~M.,  {Evans} II N.~J.,  {Lester} D.~F.,  {Levreault} R.~M.,
  {Strom} S.~E.,  1991, \apj, 376, 636

\bibitem[\protect\citeauthoryear{{Cassen} \& {Moosman}}{{Cassen} \&
  {Moosman}}{1981}]{cassen81}
{Cassen} P.,  {Moosman} A.,  1981, Icarus, 48, 353

\bibitem[\protect\citeauthoryear{{Choi}, {Kamazaki}, {Tatematsu} \&
  {Panis}}{{Choi} et~al.}{2004}]{choi04}
{Choi} M.,  {Kamazaki} T.,  {Tatematsu} K.,    {Panis} J.-F.,  2004, \apj, 617,
  1157

\bibitem[\protect\citeauthoryear{{Esquivel} \& {Raga}}{{Esquivel} \&
  {Raga}}{2007}]{esquivel07}
{Esquivel} A.,  {Raga} A.~C.,  2007, \mnras, 377, 383

\bibitem[\protect\citeauthoryear{{Foster} \& {Boss}}{{Foster} \&
  {Boss}}{1996}]{foster96}
{Foster} P.~N.,  {Boss} A.~P.,  1996, \apj, 468, 784

\bibitem[\protect\citeauthoryear{{Frank}, {King} \& {Raine}}{{Frank}
  et~al.}{2002}]{frank02}
{Frank} J.,  {King} A.,    {Raine} D.~J.,  2002, {Accretion Power in
  Astrophysics: Third Edition}.
Accretion Power in Astrophysics, by Juhan Frank and Andrew King and Derek
  Raine, pp.~398.~ISBN 0521620538.~Cambridge, UK: Cambridge University Press,
  February 2002.

\bibitem[\protect\citeauthoryear{{Hartmann}, {Ballesteros-Paredes} \&
  {Bergin}}{{Hartmann} et~al.}{2001}]{hartmann01}
{Hartmann} L.,  {Ballesteros-Paredes} J.,    {Bergin} E.~A.,  2001, \apj, 562,
  852

\bibitem[\protect\citeauthoryear{{Hartmann} \& {Burkert}}{{Hartmann} \&
  {Burkert}}{2007}]{hartmann07}
{Hartmann} L.,  {Burkert} A.,  2007, \apj, 654, 988

\bibitem[\protect\citeauthoryear{{Heitsch}, {Hartmann}, {Slyz}, {Devriendt} \&
  {Burkert}}{{Heitsch} et~al.}{2008}]{heitsch08}
{Heitsch} F.,  {Hartmann} L.~W.,  {Slyz} A.~D.,  {Devriendt} J.~E.~G.,
  {Burkert} A.,  2008, \apj, 674, 316

\bibitem[\protect\citeauthoryear{{Hennebelle}, {Whitworth}, {Cha} \&
  {Goodwin}}{{Hennebelle} et~al.}{2004}]{hennebelle04}
{Hennebelle} P.,  {Whitworth} A.~P.,  {Cha} S.-H.,    {Goodwin} S.~P.,  2004,
  \mnras, 348, 687

\bibitem[\protect\citeauthoryear{{Henriksen}, {Andre} \&
  {Bontemps}}{{Henriksen} et~al.}{1997}]{henriksen97}
{Henriksen} R.,  {Andre} P.,    {Bontemps} S.,  1997, \aap, 323, 549

\bibitem[\protect\citeauthoryear{{Huerta} \& {Mendoza}}{{Huerta} \&
  {Mendoza}}{2007}]{huerta05}
{Huerta} E.~A.,  {Mendoza} S.,  2007, Revista Mexicana de Astronomia y
  Astrofisica, 43, 191

\bibitem[\protect\citeauthoryear{{Hunter}}{{Hunter}}{1977}]{hunter77}
{Hunter} C.,  1977, \apj, 218, 834

\bibitem[\protect\citeauthoryear{{Jayawardhana}, {Hartmann} \&
  {Calvet}}{{Jayawardhana} et~al.}{2001}]{jayawardhana01}
{Jayawardhana} R.,  {Hartmann} L.,    {Calvet} N.,  2001, \apj, 548, 310

\bibitem[\protect\citeauthoryear{{Jijina}, {Myers} \& {Adams}}{{Jijina}
  et~al.}{1999}]{jijina99}
{Jijina} J.,  {Myers} P.~C.,    {Adams} F.~C.,  1999, \apjs, 125, 161

\bibitem[\protect\citeauthoryear{{Kenyon}, {Calvet} \& {Hartmann}}{{Kenyon}
  et~al.}{1993}]{kenyon93}
{Kenyon} S.~J.,  {Calvet} N.,    {Hartmann} L.,  1993, \apj, 414, 676

\bibitem[\protect\citeauthoryear{Landau \& Lifshitz}{Landau \&
  Lifshitz}{1989}]{daumech}
Landau L.,  Lifshitz E.,  1989, Mechanics, 3rd ed. edn.
Vol.~1 of Course of Theoretical Physics, Pergamon

\bibitem[\protect\citeauthoryear{{Larson}}{{Larson}}{1969}]{larson69}
{Larson} R.~B.,  1969, \mnras, 145, 271

\bibitem[\protect\citeauthoryear{{Lee} \& {Ramirez-Ruiz}}{{Lee} \&
  {Ramirez-Ruiz}}{2006}]{lee05}
{Lee} W.~H.,  {Ramirez-Ruiz} E.,  2006, \apj, 641, 961

\bibitem[\protect\citeauthoryear{{Lin} \& {Pringle}}{{Lin} \&
  {Pringle}}{1990}]{lin90}
{Lin} D.~N.~C.,  {Pringle} J.~E.,  1990, \apj, 358, 515

\bibitem[\protect\citeauthoryear{{Melioli}, {de Gouveia Dal Pino}, {de La Reza}
  \& {Raga}}{{Melioli} et~al.}{2006}]{melioli06}
{Melioli} C.,  {de Gouveia Dal Pino} E.~M.,  {de La Reza} R.,    {Raga} A.,
  2006, \mnras, 373, 811

\bibitem[\protect\citeauthoryear{{Mendoza}, {Cant{\'o}} \& {Raga}}{{Mendoza}
  et~al.}{2004}]{mendoza04}
{Mendoza} S.,  {Cant{\'o}} J.,    {Raga} A.~C.,  2004, Revista Mexicana de
  Astronomia y Astrofisica, 40, 147

\bibitem[\protect\citeauthoryear{{Michel}}{{Michel}}{1972}]{michel72}
{Michel} F.~C.,  1972, \apss, 15, 153

\bibitem[\protect\citeauthoryear{{Motte} \& {Andr{\'e}}}{{Motte} \&
  {Andr{\'e}}}{2001}]{motte01}
{Motte} F.,  {Andr{\'e}} P.,  2001, \aap, 365, 440

\bibitem[\protect\citeauthoryear{{Mouschovias} \& {Paleologou}}{{Mouschovias}
  \& {Paleologou}}{1979}]{mouschovias79}
{Mouschovias} T.~C.,  {Paleologou} E.~V.,  1979, \apj, 230, 204

\bibitem[\protect\citeauthoryear{{Nagel}}{{Nagel}}{2007}]{nagel07}
{Nagel} E.,  2007, Revista Mexicana de Astronomia y Astrofisica, 43, 257

\bibitem[\protect\citeauthoryear{{Penston}}{{Penston}}{1969}]{penston69}
{Penston} M.~V.,  1969, \mnras, 144, 425

\bibitem[\protect\citeauthoryear{{Shu}}{{Shu}}{1977}]{shu77}
{Shu} F.~H.,  1977, \apj, 214, 488

\bibitem[\protect\citeauthoryear{{Shu}, {Adams} \& {Lizano}}{{Shu}
  et~al.}{1987}]{shu87}
{Shu} F.~H.,  {Adams} F.~C.,    {Lizano} S.,  1987, \araa, 25, 23

\bibitem[\protect\citeauthoryear{{Stahler}, {Korycansky}, {Brothers} \&
  {Touma}}{{Stahler} et~al.}{1994}]{stahler94}
{Stahler} S.~W.,  {Korycansky} D.~G.,  {Brothers} M.~J.,    {Touma} J.,  1994,
  \apj, 431, 341

\bibitem[\protect\citeauthoryear{{Throop} \& {Bally}}{{Throop} \&
  {Bally}}{2008}]{throop08}
{Throop} H.~B.,  {Bally} J.,  2008, \aj, 135, 2380

\bibitem[\protect\citeauthoryear{{Ulrich}}{{Ulrich}}{1976}]{ulrich76}
{Ulrich} R.~K.,  1976, \apj, 210, 377

\bibitem[\protect\citeauthoryear{{V{\'a}zquez-Semadeni}, {G{\'o}mez},
  {Jappsen}, {Ballesteros-Paredes}, {Gonz{\'a}lez} \&
  {Klessen}}{{V{\'a}zquez-Semadeni} et~al.}{2007}]{vazquez07}
{V{\'a}zquez-Semadeni} E.,  {G{\'o}mez} G.~C.,  {Jappsen} A.~K.,
  {Ballesteros-Paredes} J.,  {Gonz{\'a}lez} R.~F.,    {Klessen} R.~S.,  2007,
  \apj, 657, 870

\bibitem[\protect\citeauthoryear{{Ward-Thompson}, {Motte} \&
  {Andre}}{{Ward-Thompson} et~al.}{1999}]{ward-thompson99}
{Ward-Thompson} D.,  {Motte} F.,    {Andre} P.,  1999, \mnras, 305, 143

\bibitem[\protect\citeauthoryear{{Whitney}, {Wood}, {Bjorkman} \&
  {Wolff}}{{Whitney} et~al.}{2003}]{whitney03}
{Whitney} B.~A.,  {Wood} K.,  {Bjorkman} J.~E.,    {Wolff} M.~J.,  2003, \apj,
  591, 1049

\bibitem[\protect\citeauthoryear{{Whitworth} \& {Ward-Thompson}}{{Whitworth} \&
  {Ward-Thompson}}{2001}]{whitworth01}
{Whitworth} A.~P.,  {Ward-Thompson} D.,  2001, \apj, 547, 317

\end{thebibliography}

\label{lastpage}

\end{document}